\documentclass[a4paper]
{cedram-smai-jcm}

\usepackage{subcaption}
\usepackage{hyperref}
\usepackage{graphicx}
\usepackage{amssymb}
\usepackage{color}
\usepackage{soul}

\graphicspath{ {./Figures/} }

\title[]{A robust, discrete-gradient descent procedure for optimisation with time-dependent PDE and norm constraints}

\author{PM Mannix}
\address{Universit\'e C\^ote d’Azur, Inria, CNRS, LJAD, France} 
\email{Paul.Mannix@univ-cotedazur.fr}

\author{CS Skene}
\address{Department of Applied Mathematics, University of Leeds, West Yorkshire, UK}

\author{D Auroux}\address{Universit\'e C\^ote d’Azur, Inria, CNRS, LJAD, France} 
\author{F Marcotte}\address{Universit\'e C\^ote d’Azur, Inria, CNRS, LJAD, France} 

  

\begin{document}

\begin{abstract}
Many physical questions in fluid dynamics can be recast in terms of norm constrained optimisation problems; which in-turn, can be further recast as unconstrained problems on spherical manifolds. Due to the nonlinearities of the governing PDEs, and the computational cost of performing optimal control on such systems, improving the numerical convergence of the optimisation procedure is crucial. Borrowing tools from the optimisation on manifolds community we outline a numerically consistent, discrete formulation of the direct-adjoint looping method accompanied by gradient descent and line-search algorithms with global convergence guarantees. We numerically demonstrate the robustness of this formulation on three example problems of relevance in fluid dynamics and provide an accompanying library \texttt{SphereManOpt}.
\end{abstract}

\maketitle

\section{Introduction}

Applications of norm constrained, nonlinear optimal control abound in fluid dynamics: spanning from the improvement of transport and mixing efficiencies in complex fluids (e.g. \cite{foures2013localization,eggl2018gradient}), to the stability analysis of thermo-acoustic oscillations (e.g. \cite{juniper2018sensitivity}), the modeling of transitions between a dynamical system's multiple basins of attraction (e.g. \cite{Lecoanet_SHE, Kerswell2018}) or the forecasts of oceanographic systems (e.g. \cite{terwisscha2008conditional,mu2010extension,auroux2005data}). Because the governing partial differential equations (PDEs) are fully nonlinear, solutions to the corresponding optimisation problems must in general be calculated numerically. Due to the large dimension of both the control and state vectors and the significant cost of direct numerical simulations, efficient optimisation routines must be applied to address such problems. As our motivation is essentially numerical, we will restrict our attention to finite-dimension problems in what follows.

Given a discretised set of equations describing the evolution of the state vector $\boldsymbol{X} \in \mathbb{R}^n$, a general formulation of such norm-constrained problem is as follows
\begin{equation}
\begin{split}
\min_{ \boldsymbol{X}_0} \quad & J(\boldsymbol{X},\boldsymbol{X}_0), \\
\hbox{s.t.} \quad & \boldsymbol{G}(\boldsymbol{X},\boldsymbol{X_0}) \equiv 
\begin{cases}
\frac{d \boldsymbol{X}}{d t} - \boldsymbol{F}(\boldsymbol{X},t) = 0,\\
\boldsymbol{X}_0 - \boldsymbol{X}(t=0) =0, 
\end{cases} \\
& c(\boldsymbol{X}_0) \equiv || \boldsymbol{X}_0 ||^2 - E_0 = 0,  
\end{split}
\label{eq:Optimisation_Problem_General}
\end{equation}
where $\boldsymbol{X}_0 \in \mathbb{R}^n$ is the control variable we seek to optimise, and $E_0$ its energy. The objective function $J(\boldsymbol{X},\boldsymbol{X}_0)$ is a quantity of interest to be minimised (or maximised), and $\boldsymbol{F}(\boldsymbol{X},t)$ is a nonlinear operator describing how the system evolves in time. For example, $\boldsymbol{X}_0$ can be the initial velocity field of kinetic energy $E_0$, yielding optimal mixing over a finite time window $t \in (0,T]$ as the flow evolves according to its governing equations. In this case, well-mixedness can be achieved by minimising some measure of the fluid heterogeneity (such as the scale-sensitive mix-norms \cite{mathew2005multiscale,thiffeault2012using,heffernan2022robust}). In the context of stability analysis, $\boldsymbol{X}_0$ can be the initial perturbation, which, for a given energy input, maximises for example the nonlinear growth of perturbations - thus allowing for a better characterisation of laminar-turbulent transitions in shear flows \cite{Pringle2010,Kerswell2018}. Although \eqref{eq:Optimisation_Problem_General} assumes control of the initial condition only, this can be easily modified to consider a time-dependent forcing or an alternative control variable.

\subsection{Adjoint formalism}\label{Adjoint_Motivation}

To solve the optimisation problem \eqref{eq:Optimisation_Problem_General} one formulates the Lagrangian $\mathcal{L}$, related to the pure objective functional, as
\begin{equation}
\mathcal{L}(\boldsymbol{X},\boldsymbol{X}_0; \boldsymbol{q}, \lambda) \equiv \hat{J}(\boldsymbol{X}_0) + \lambda c(\boldsymbol{X_0}) = J(\boldsymbol{X},\boldsymbol{X}_0) + \int_0^T \langle \boldsymbol{q}, \boldsymbol{G}(\boldsymbol{X},\boldsymbol{X_0})\rangle \; dt + \lambda( || \boldsymbol{X}_0 ||^2  - E_0),
\label{eq:Lagrangian_Functional}
\end{equation}
where $\boldsymbol{q},\lambda$ denote the Lagrange multipliers (also termed adjoint variables) used to enforce these constraints, $\langle \boldsymbol{a}, \boldsymbol{b} \rangle$ denotes a vector inner product and $|| \boldsymbol{a} ||^2 = \langle \boldsymbol{a}, \boldsymbol{a} \rangle$ the associated norm. Here we choose to denote the pure objective functional by $\hat{J}(\boldsymbol{X}_0)$, thus assuming $\boldsymbol{X}(\boldsymbol{X}_0)$ is implicitly defined for all $\boldsymbol{X}_0$ by the constraint equations. Any stationary point of $\mathcal{L}$ satisfies the following Euler-Lagrange equations:
\begin{eqnarray} 
\frac{\delta \mathcal{L}}{\delta \lambda} = || \boldsymbol{X}_0 ||^2 - E_0 = 0,
\label{eq:Euler_Lagrangian_Eqns_1} \\    
\frac{\delta \mathcal{L}}{\delta \boldsymbol{q}} = \boldsymbol{G}(\boldsymbol{X},\boldsymbol{X_0}) = 0, \\
\frac{\delta \mathcal{L}}{\delta \boldsymbol{X}_T} = \frac{\delta J}{\delta \boldsymbol{X}_T} + \boldsymbol{q}_T = 0, \label{eq:Euler_Lagrangian_Eqns_Compatibility} \\
\frac{\delta \mathcal{L}}{\delta \boldsymbol{X} } = \frac{\partial \boldsymbol{G}}{\partial \boldsymbol{X}}^{\dagger} \boldsymbol{q} + \frac{\delta J}{\delta \boldsymbol{X}} = 0, \label{eq:Euler_Lagrangian_Eqns_Adjoint} \\
\frac{\delta \mathcal{L}}{\delta \boldsymbol{X}_0} = \frac{\delta J}{\delta \boldsymbol{X}_0} - \boldsymbol{q}_0 + 2 \lambda \boldsymbol{X}_0 =0,
\label{eq:Euler_Lagrangian_Eqns}
\end{eqnarray}
where $\boldsymbol{X}_T,\boldsymbol{q}_T$ refers to the vectors at the final time, while $\dagger$ notation is used to denote the adjoint operator such that $\langle x,Ly \rangle=\langle L^\dagger x,y \rangle$. In general, reaching a stationary point of \eqref{eq:Lagrangian_Functional} demands an iterative procedure, whereby we solve \eqref{eq:Euler_Lagrangian_Eqns_1} to \eqref{eq:Euler_Lagrangian_Eqns} from top to bottom as given. That is: a random initial vector $\boldsymbol{X}_0$ is normalised to be of norm $E_0$, and the PDE constraint equations are solved forward in time from $t=0 \to T$ using an appropriate time-stepping scheme \cite{Ascher1995, Ascher1997}, whilst simultaneously calculating $\hat{J}(\boldsymbol{X}_0)$. Subject to the final time condition \eqref{eq:Euler_Lagrangian_Eqns_Compatibility}, the adjoint equations \eqref{eq:Euler_Lagrangian_Eqns_Adjoint}
\begin{equation}
\big( \frac{\partial }{\partial t} + \frac{\partial \boldsymbol{F}}{\partial \boldsymbol{X}}^{\dag} \big)\boldsymbol{q} = \frac{\delta J}{\delta \boldsymbol{X}},
\label{eq:adjoint_equation}
\end{equation}
are then solved backwards from $t=T \to 0$. Denoting the current and subsequent iterates of a gradient descent procedure by $k,k+1$ respectively, the control parameter is finally updated, using the gradient at iteration $k$ \eqref{eq:Euler_Lagrangian_Eqns_Compatibility}, via
\begin{equation}
\boldsymbol{X}^{k+1}_0 = \boldsymbol{X}^k_0 - \alpha \frac{\delta \mathcal{L}}{\delta \boldsymbol{X}_0     }^k(\lambda),
\label{eq:norm_lagrange_update}
\end{equation}
where $\alpha \in \mathbb{R}^+$ is an appropriately chosen step-size to ensure descent \cite{wright1999numerical} and $\lambda$ is determined to enforce $|| \boldsymbol{X}^{k+1}||^2 = E_0$.

The first challenge when performing gradient descent is to accurately and efficiently estimate the gradient $\frac{\delta \mathcal{L}}{\delta \boldsymbol{X}_0}^k(\lambda)$. Using open-source automatic differentiation (AD) to obtain discrete gradients is one possible solution \cite{TapenadeRef13}. While this approach works well on forward codes written so as to be amenable to AD, such as dolfin-adjoint for example \cite{farrell2013automated,Mitusch2019}, it can otherwise consume large amounts of memory and be extremely challenging to debug \cite{muller2005performance,naumann2011art}. Moreover, using AD to differentiate MPI-parallel codes also raises important difficulties regarding how to manage communication dataflow and dependencies \cite{utke2009toward,maugars2022algorithmic}, which in practice hinders its applicability to computationally expensive fluid dynamics problems. When impractical to use AD for some of the reasons mentioned above, the alternative approach is to derive analytically the discrete adjoint equations using summation by parts. This approach, albeit established in the optimisation literature \cite{Hager2000,Zhang2022tsadjoint}, has been less discussed until recently in the aeronautics \cite{Zahr2016,Franco2019} and fluid dynamics literature \cite{vishnampet2015practical, schmid2017stability, kord2019optimal, fikl2020control}.

Even when armed with an accurate gradient however, we must move in the correct direction and use an efficient gradient descent routine. Typically one enforces the equality constraint $c(\boldsymbol{X}_0)$ using a Lagrange multiplier $\lambda$ as outlined above. Given however that this constraint naturally restricts $\boldsymbol{X}_0 \in \mathbb{R}^n$ to the spherical manifold $\mathcal{S}^{n-1} = \{ \boldsymbol{X}_0 \; | \; || \boldsymbol{X}_0 ||^2 = E_0 \}$, we can implement a consistent and therefore more efficient routine by exploiting this structure \cite{absil2009optimization,sato2021riemannian_book}, which we recall in \autoref{sec:Linesearch_Algorithim}.

\subsection{Issue 1: Discrete versus continuous}\label{sec:Iissue_1}

To highlight the inconsistencies that typically arise when discretizing the previous (continuous) derivation, we recall that, although a linear operator $L = \partial_{xx}$ acting on function $u(x), -1 \leq x \leq 1$ with boundary data $u(\pm 1) =0$ is self-adjoint with respect to $\int_{-1}^1 p L u \, dx = \int_{-1}^1 L p u \, dx,$ its corresponding matrix operator although dependent on the discretization used $L^{\dagger}_{i,j} \neq L_{i,j}$ seldom conserves this property \cite{hinze2008optimization,leykekhman2012investigation}. Similarly, integration by parts of the temporal derivative followed by discretisation, does not generally commute with discretisation followed by discrete integration by parts. To clarify this issue we consider the following discrete Lagrangian consisting of a linear constraint equation $\boldsymbol{X}_t = L\boldsymbol{X}$, discretised using a Crank-Nicolson temporal scheme
\begin{equation}
    \mathcal{L}(\boldsymbol{X},\boldsymbol{X}_0; \boldsymbol{q}, \lambda) = J({\boldsymbol{X}},\boldsymbol{X}_0) - \langle \boldsymbol{q}, M \boldsymbol{{X}} - \boldsymbol{f} \rangle + \lambda( ||\boldsymbol{X}_0 ||^2  - E_0),
    \label{eq:Disc_Lagrangian_Functional}
\end{equation}
where $\Delta t = T/N, \boldsymbol{{X}} = \left( \boldsymbol{X}_0, \cdots, \boldsymbol{X}_N \right)$ and the constraint equations $M \boldsymbol{{X}} = \boldsymbol{f}$ (incorporating the constraint on the initial condition) now in discrete form read
\begin{equation}
    \left( 
    {\begin{array}{*{5}c}
    I & 0 &  \cdots &  &  \\
    b                  & a &  \ddots &  &  \\
    0                  & \ddots & \ddots & \ddots & \vdots \\
    \vdots             & \ddots & \ddots & \ddots & 0  \\
                       & \cdots & 0      & b & a  \\
    \end{array}} 
    \right)
    \left( 
    {\begin{array}{c}
    \boldsymbol{X}_0 \\ \boldsymbol{X}_1 \\ \vdots \\ \vdots \\ \boldsymbol{X}_N
    \end{array}} 
    \right) =
    \left( 
    {\begin{array}{c}
    \boldsymbol{X}(t=0) \\ 0 \\ \vdots \\ \vdots \\ 0
    \end{array}}
    \right),
\label{eq:Vars_Disc_Lagrangian_Functional_1}
\end{equation}
where $a,b = (\frac{I}{\Delta t} - \frac{L}{2}),-(\frac{I}{\Delta t} + \frac{L}{2})$. Taking variations with respect to $\boldsymbol{X}$ we obtain $M^{\dagger} \boldsymbol{{q}} = \frac{\delta J}{\delta {\boldsymbol{X}}} + 2 \lambda {\boldsymbol{X}_0}$ which in component form reads
\begin{equation}
    \left(
    \begin{array}{*{5}c}
    I & b^{\dag} &  0 & \cdots  &  \\
    0 & a^{\dag} & \ddots &  & \vdots  \\
    \vdots & \ddots & \ddots & \ddots & 0 \\
     &  & \ddots & \ddots & b^{\dag}  \\
     &  & \cdots & 0 & a^{\dag}  \\
    \end{array}
    \right)
    \left(
    \begin{array}{c}
    \boldsymbol{q}_0 \\ \boldsymbol{q}_1 \\ \vdots \\ \vdots \\ \boldsymbol{q}_{N}
    \end{array} \right)
    = 
    \left(
    \begin{array}{c}
     \frac{\delta J}{\delta \boldsymbol{X}_0} + 2 \lambda \boldsymbol{X}_0 \\ \frac{\delta J}{\delta \boldsymbol{X}_1} \\ \vdots \\ \vdots \\ \frac{\delta J}{\delta \boldsymbol{X}_N}
    \end{array} 
    \right).
\label{eq:Vars_Disc_Lagrangian_Functional_2}
\end{equation}
Comparing these discrete Euler-Lagrange equations \eqref{eq:Vars_Disc_Lagrangian_Functional_2} with their continuous counterpart \eqref{eq:Euler_Lagrangian_Eqns_Compatibility}-\eqref{eq:Euler_Lagrangian_Eqns} one notices that: (1) the adjoint reverses the direction of information propagation, (2) the discrete adjoint of a \emph{centered} differencing scheme (here for the temporal derivative) would remain consistent with its continuous adjoint if and only if $L = L^{\dag}$ and (3) the discrete terminal/compatibility conditions at $t=0,T$ differ from their continuous counterpart at $n=0,N$. As we will show in this paper, the later issue is compounded for higher order schemes which contribute additional variations to more of the top and bottom rows. Throughout the remainder of this paper the term \emph{discrete gradient} will be used to refer to the approach where the adjoint numerical scheme is obtained by differentiating the already-discretized direct equations. Conversely \emph{continuous gradient} will be used here to refer to the approach where the adjoint equations are first derived from the continuous problem, then discretized. A scheme for which the solutions to both approaches coincide is termed adjoint/dual consistent \cite{hartmann2007adjoint,BERG201341}.

\subsection{Issue 2: Constrained gradient update}\label{sec:issue_2}

The second issue is that we still have a norm-constrained update step \eqref{eq:norm_lagrange_update}. For every choice of step-size $\alpha$ we must apply the normalisation constraint to choose a corresponding $\lambda$. This relegates the applicability of line-search algorithms for unconstrained optimisation, which allow us to choose step-sizes $\alpha$ guaranteeing descent \cite{wright1999numerical}. Aside from the in-applicability of standard line-search algorithms however, further issues exist with the Lagrange multiplier approach. Using $\nabla f = \frac{\delta f}{\delta \boldsymbol{X}_0}$ to denote the variational derivative, it follows that for an optimum to exist, a necessary condition is that a $\lambda$ exists such that
\begin{equation}
\frac{\delta \mathcal{L}}{\delta \boldsymbol{X}_0     } = \nabla \hat{J}(\boldsymbol{X}_0) + \lambda \nabla c(\boldsymbol{X}_0) = 0,
\label{eq:necessary_condition}
\end{equation}
implying that the gradient of the constraint equations $\nabla \hat{J}(\boldsymbol{X}_0)$ and norm constraint $\nabla c(\boldsymbol{X}_0)$ are parallel. To render this condition sufficient requires that $\lambda$ satisfy the Karush–Kuhn–Tucker (KKT) conditions \cite{wright1999numerical}, which in-turn requires computing the Hessian of \eqref{eq:necessary_condition}. If $\nabla \hat{J}(\boldsymbol{X}_0), \nabla c(\boldsymbol{X}_0)$ are not parallel, \eqref{eq:necessary_condition} is not satisfied and we ideally, according to a first order approximation, choose a search direction $\boldsymbol{d}$ such that $\nabla \hat{J}(\boldsymbol{X}_0)^T\boldsymbol{d} < 0$ and $\nabla c(\boldsymbol{X}_0)^T\boldsymbol{d} = 0$. A good candidate which also defines the gradient on the spherical manifold $\mathcal{S}$ \cite{boumal2020introduction}, is the tangent-vector 
\begin{equation}
-\boldsymbol{d} = \nabla \hat{J}(\boldsymbol{X}_0) - \bigg( \frac{ \nabla c(\boldsymbol{X}_0)^T \nabla \hat{J}(\boldsymbol{X}_0) }{||\nabla c(\boldsymbol{X}_0)||^2} \bigg) \nabla c(\boldsymbol{X}_0),
\label{eq:tangent_vector}
\end{equation}
as it satisfies the previous inequality and equality constraints but crucially is independent of $\lambda$. Thus by restricting $\boldsymbol{X}_0$ to $\mathcal{S}$ and updating $\boldsymbol{X}_0$ on geodesics of $\mathcal{S}$ for all values of the step size $\alpha$ \cite{Douglas1998}, we can omit the equality constraint in \eqref{eq:Lagrangian_Functional}. Furthermore $\nabla \hat{J}(\boldsymbol{X}_0) = 0$ now provides a necessary condition for a local minimiser \cite{boumal2020introduction}. 

What remains, following \cite{sato2021riemannian_article,sato2021riemannian_book}, is to appropriately adapt the existing line-search algorithms such that updating $\boldsymbol{X}_0^k$ leads to global convergence, and to assess their efficiency on a suite of non-trivial fluid dynamics examples. Aside from the \texttt{Pymanopt} library for optimisation on matrix manifolds \cite{JMLR:v17:16-177}, SLSQP \cite{kraft1988software} presents the only alternative for problems with equality constraints. Owing to the construction, storage $\mathcal{O}(n^2)$ and inversion $\mathcal{O}(n^3)$ of the Hessian matrix SLSQP is however numerically too demanding to handle high-dimension control variables \cite{wright1999numerical, johnson2014nlopt}. Conversely while \texttt{Pymanopt} implements computationally efficient methods, its linesearch does not satisfy the necessary conditions in order to guarantee global convergence \cite{sato2021riemannian_article}.

In what follows, we address the previously highlighted issues and test their numerical remedy against a suite of physical problems. Given the additional complexity represented by the derivation of a consistent, discrete adjoint numerical scheme, compared to a continuous gradient approach (where the same numerical scheme is used to discretize the forward and backward equations), we also conduct a systematic numerical comparison between discrete and continuous adjoint optimisation, which excluding the studies of \cite{vishnampet2015practical} and \cite{fikl2016comprehensive} is currently lacking. To this end, we first derive the discrete adjoint equations in \autoref{sec:Discrete_Adjoint}; following \cite{sato2021riemannian_book} we then outline a consistent gradient update with guaranteed global descent in \autoref{sec:Linesearch_Algorithim}; and finally we combine these developments in \autoref{sec:Numerical_Comparisons} to demonstrate their superior performance. Perspectives for further work are then discussed in the conclusive section.

\section{Discrete adjoint equations}\label{sec:Discrete_Adjoint}

A set of \emph{spatially discretised} coupled evolution equations can be written as
\begin{equation}
    \mathcal{M} \partial_t \boldsymbol{X} + \mathcal{L} \boldsymbol{X} = \mathcal{F}(\boldsymbol{X},t),
    \label{eq:TimeStep_Notation}
\end{equation}
where $\boldsymbol{X}$ is the vector of unknowns, $\mathcal{M},\mathcal{L}$ are linear operators encompassing both the constraint equations and system boundary conditions and the right hand side $\mathcal{F}$ encompasses all nonlinear and non-autonomous terms. Then we can write a general \emph{s-step} multistep implicit-explicit (IMEX) method \cite{Ascher1995} as 
\begin{equation}
    \sum_{j=0}^s a_j \mathcal{M} \boldsymbol{X}_{n-j} + \sum_{j=0}^s b_j \mathcal{L} \boldsymbol{X}_{n-j} = \sum_{j=1}^s c_j \mathcal{F}_{n-j},
    \label{eq:TimeStep_Notation_multistep}
\end{equation}
where index $n$ denotes the $n^{th}$ time step, $j$ the internal stages and the right hand side is evaluated and treated as a vector. Notably when initiating time-stepping from a single initial condition, we have insufficient data to use 2nd or higher order schemes directly and must revert to a first order scheme. While diagonally implicit Runge-Kutta (DIRK/SDIRK) integration schemes \cite{Ascher1997} solve this issue and are L-stable, their numerical overhead, both in terms of CPU and memory, is much higher however and their discrete adjoint derived by \cite{Hager2000,Sandu2006,vishnampet2015practical,Zahr2016} complexifies faster.

We first present the discrete adjoint equations in their full generality, and subsequently consider their numerical performance on particular examples by making comparisons with their continuous formulation. As multistep schemes of order $s > 2$ are not A-stable and objective functions can be time-integrated with $\mathcal{O}(\Delta t^2)$ accuracy at best, higher order schemes are not considered. We note that the adjoint of a subclass of IMEX schemes known as backward differencing schemes (BDF) has previously been derived up to third order by \cite{nielsen2013discrete}.

\subsection{A second order IMEX scheme }

Partitioning the time interval $t \in (0,T]$ as $t_n = n \Delta t, n \in (0,N=T/\Delta t]$ the discrete Lagrangian, with normalisation constraints omitted, is given as 
\begin{equation}
\begin{split}
 \mathcal{L} &= J(\boldsymbol{X}) \\
& - \langle \boldsymbol{q}_0, \boldsymbol{X}_0 - \boldsymbol{X}(\boldsymbol{x},t=0) \rangle \\
& -\langle \boldsymbol{q}_1, \mathcal{M}( \tilde{a}_0 \boldsymbol{X}_1 + \tilde{a}_1 \boldsymbol{X}_0) + \mathcal{L}( \tilde{b}_0 \boldsymbol{X}_1 + \tilde{b}_1 \boldsymbol{X}_0) - \tilde{c}_1 \mathcal{F}(\boldsymbol{X}_0) \rangle \\
& -\sum_{n=1}^{N-1} \langle \boldsymbol{q}_{n+1}, \mathcal{M}(\tilde{a}_0 \boldsymbol{X}_{n+1} + \tilde{a}_1 \boldsymbol{X}_n + a_2 \boldsymbol{X}_{n-1}) + \mathcal{L}(\tilde{b}_0 \boldsymbol{X}_{n+1} + \tilde{b}_1 \boldsymbol{X}_n + b_2 \boldsymbol{X}_{n-1})  \\
& - \bigg[ \tilde{c}_1 \mathcal{F}(\boldsymbol{X}_n) + c_2 \mathcal{F}(\boldsymbol{X}_{n-1}) \bigg] \rangle ,
\end{split}
\label{eq:Max_Lagrangian_Discrete}
\end{equation}
where tilde notation is used to denote the coefficients retained if considering a first order scheme only, and the objective function is for example given by either
\begin{equation}
    J(\boldsymbol{X}) = \frac{\Delta t}{2} \big( \langle \boldsymbol{X}_0, W\boldsymbol{X}_0 \rangle + \langle \boldsymbol{X}_N, W\boldsymbol{X}_N \rangle \big)  + \Delta t \sum_{n=1}^{N-1} \langle \boldsymbol{X}_n, W\boldsymbol{X}_n \rangle,
    \label{eq:Disc_Objective_App}
\end{equation}
or $J_T(\boldsymbol{X}) = \langle \boldsymbol{X}_N, W \boldsymbol{X}_N \rangle$ where $W$ is a known matrix operator. Taking variations with respect to the adjoint variables $\boldsymbol{q}_n$ and setting $\frac{\delta \mathcal{L}}{\delta \boldsymbol{q}_n} = 0$ we recover the initial condition and the constraint equations, while with respect to the primal variables $\boldsymbol{X}_n$ we obtain
\begin{equation}
\begin{split}
\sum_{n=0}^N \langle \delta \boldsymbol{X}_n, \frac{ \delta \mathcal{L} }{\delta \boldsymbol{X}_n }  \rangle  &=  \sum_{n=0}^N \langle \delta \boldsymbol{X}_n, \frac{\partial J}{\partial \boldsymbol{X}_n} \rangle  -\langle \delta \boldsymbol{X}_0,  \boldsymbol{q}_0 \rangle \\
& -\langle \delta \boldsymbol{X}_1, \tilde{a}_0 \mathcal{M}^{\dagger} \boldsymbol{q}_1 + \tilde{b}_0 \mathcal{L}^{\dagger} \boldsymbol{q}_1 \rangle - \langle \delta \boldsymbol{X}_0, \tilde{a}_1 \mathcal{M}^{\dagger} \boldsymbol{q}_1 + \tilde{b}_1 \mathcal{L}^{\dagger} \boldsymbol{q}_1 -\tilde{c}_1 \frac{ \partial F }{\partial \boldsymbol{X}_0}^{\dagger} \boldsymbol{q}_1 \rangle \\
& -\sum_{n=2}^{N} \langle \delta \boldsymbol{X}_n, \tilde{a}_0 \mathcal{M}^{\dagger} \boldsymbol{q}_n + \tilde{b}_0 \mathcal{L}^{\dagger} \boldsymbol{q}_n \rangle \\
& -\sum_{n=1}^{N-1} \langle \delta \boldsymbol{X}_n, \tilde{a}_1 \mathcal{M}^{\dagger} \boldsymbol{q}_{n+1} + \tilde{b}_1 \mathcal{L}^{\dagger} \boldsymbol{q}_{n+1} \rangle \\
& -\sum_{n=0}^{N-2} \langle \delta \boldsymbol{X}_n, a_2 \mathcal{M}^{\dagger} \boldsymbol{q}_{n+2} + b_2 \mathcal{L}^{\dagger} \boldsymbol{q}_{n+2} \rangle \\
& -\sum_{n=1}^{N-1} \langle \delta \boldsymbol{X}_{n},-\tilde{c}_1  \frac{ \partial F }{\partial \boldsymbol{X}_n}^{\dagger} \boldsymbol{q}_{n+1} \rangle -\sum_{n=0}^{N-2} \langle \delta \boldsymbol{X}_n,-c_2 \frac{ \partial F }{\partial \boldsymbol{X}_n}^{\dagger} \boldsymbol{q}_{n+2} \rangle,
\end{split}
\label{eq:Max_Lagrangian_Var1}    
\end{equation}
where we have permuted the indices in each of these terms and $\frac{ \partial F }{\partial \boldsymbol{X}_n}$ is used to denote the Jacobian of the nonlinear term. Grouping terms of \eqref{eq:Max_Lagrangian_Var1} by their indices and setting them equal to zero we obtain the discrete Euler-Lagrange equations for $n=N,N-1$, $N-2 \geq n \geq 1$ and $n=0$ respectively
\begin{eqnarray}
\big( \tilde{a}_0 \mathcal{M}^{\dagger}  + \tilde{b}_0 \mathcal{L}^{\dagger} \big) \boldsymbol{q}_N = \frac{\partial J}{\partial \boldsymbol{X}_N}, \label{eq:Disc_Euler_Lagrange_CNAB1_0}
\\
\big( \tilde{a}_0 \mathcal{M}^{\dagger} + \tilde{b}_0 \mathcal{L}^{\dagger} \big) \boldsymbol{q}_{N-1} + \big( \tilde{a}_1 \mathcal{M}^{\dagger} + \tilde{b}_1 \mathcal{L}^{\dagger} \big) \boldsymbol{q}_N - \tilde{c}_1 \frac{ \partial F }{\partial \boldsymbol{X}_{N-1}}^{\dagger} \boldsymbol{q}_N = \frac{\partial J}{\partial \boldsymbol{X}_{N-1}},
\\
\big( \tilde{a}_0 \mathcal{M}^{\dagger} + \tilde{b}_0 \mathcal{L}^{\dagger} \big) \boldsymbol{q}_{n} + \big( \tilde{a}_1 \mathcal{M}^{\dagger} + \tilde{b}_1 \mathcal{L}^{\dagger} \big) \boldsymbol{q}_{n+1}  + \big( a_2 \mathcal{M}^{\dagger} + b_2 \mathcal{L}^{\dagger} \big) \boldsymbol{q}_{n+2} - \frac{ \partial F }{\partial \boldsymbol{X}_n}^{\dagger} \big[ \tilde{c}_1 \boldsymbol{q}_{n+1} + c_2 \boldsymbol{q}_{n+2} \big] = \frac{\partial J}{\partial \boldsymbol{X}_n},
\\
\boldsymbol{q}_0 + \big( \tilde{a}_1 \mathcal{M}^{\dagger} + \tilde{b}_1 \mathcal{L}^{\dagger} \big) \boldsymbol{q}_1 + \big( a_2 \mathcal{M}^{\dagger} + b_2 \mathcal{L}^{\dagger} \big) \boldsymbol{q}_2 - \frac{ \partial F }{\partial \boldsymbol{X}_0}^{\dagger} \big[ \tilde{c}_1 \boldsymbol{q}_1 + c_2 \boldsymbol{q}_2 \big] = \frac{\partial J}{\partial \boldsymbol{X}_0}.
\label{eq:Disc_Euler_Lagrange_CNAB1}
\end{eqnarray}
The solution of the $\mathcal{O}(\Delta t)$ accurate counterpart of these equations using a preconditioned scheme, applicable for example to discretisations of wall bounded domains is given in Appendix \ref{App:Preconditioned_First_Order_Scheme}. Concentrating on the minor and feasible modifications of simple schemes which help achieve accurate convergence, we have implemented primarily first order schemes in this paper. 

Comparing expressions \eqref{eq:Disc_Euler_Lagrange_CNAB1_0}-\eqref{eq:Disc_Euler_Lagrange_CNAB1} with their continuous counterpart \eqref{eq:Euler_Lagrangian_Eqns_Compatibility}-\eqref{eq:Euler_Lagrangian_Eqns}, we see that while the time-stepping scheme is unaffected for $s \leq n \leq N-1$, the compatibility conditions demand the inversion of non-local linear operators at $n=N,t=T$ and an explicit evaluation at $n=0,t=0$. For a fully periodic domain discretised using a self-adjoint Fourier basis the error is likely to be small, but for wall bounded domains where boundary layers are present, such as the optimal mixing example presented in \ref{sec:Optimal_Mixing_Grad_Test}, the error is potentially significant.

\subsection{Numerical performance}\label{sec:Test_cases}

To validate the adjoint of the discrete equations and justify the additional complexity of its implementation we now report numerical experiments on three example cases that showcase its superior performance. The consistency of the adjoint is checked by testing the Taylor remainder, i.e. verifying that for any arbitrary perturbation $\delta \boldsymbol{X}$ of the initial condition,
\begin{equation}
    \lim_{ h \to 0 } \; | \hat{J}(\boldsymbol{X}_0 + h \delta \boldsymbol{X} ) - \hat{J}(\boldsymbol{X}_0) | \to 0, \quad \hbox{at} \quad \mathcal{O}(|h|),
\label{eq:Grad_est}
\end{equation}
and
\begin{equation}
    \lim_{ h \to 0 } \; | \hat{J}(\boldsymbol{X}_0 + h \delta \boldsymbol{X} ) - \hat{J}(\boldsymbol{X}_0) - h \langle \delta \boldsymbol{X}, \frac{\delta \mathcal{L}}{\delta \boldsymbol{X}_0} \rangle | \to 0, \quad \hbox{at} \quad \mathcal{O}(|h|^2).
    \label{eq:Tay_R2_test}
\end{equation}
An additional test of interest is that the first difference in \eqref{eq:Tay_R2_test}, divided by $h$, is equivalent to the inner product in the last term, to a number of decimal places. This is preferable when $h \ll 1$ as it avoids multiplying very small and large numbers. In all tests which follow we generate $\boldsymbol{X}_0, \delta \boldsymbol{X}_0$ randomly and re-scale such that $||\boldsymbol{X}_0||^2=||\delta \boldsymbol{X}_0||^2=1$. In practice, it is strongly advised whenever possible to also choose $\boldsymbol{X}_0$ close to a representative state of interest of the problem studied, as this ensures consistency of gradient definition with respect to problem relevant parameter vectors.

\subsubsection{First test-case: a dynamo problem}\label{sec:Ex_Kinematic_Dynamo}

In many astrophysical objects, magnetic fields are generated by the motions of electrically conducting fluids, through a process known as the dynamo effect \cite{larmor201317}. A dynamo optimisation problem first considered by \cite{Willis2012}, comprises finding the velocity field $\boldsymbol{U}$ and the initial magnetic field $\boldsymbol{B}_0$ (constrained by some energy budget), which maximise the magnetic field's growth by some target time $T$ in a triply periodic box. Here the velocity field is assumed to be frozen, i.e. independent of time and the evolution of the magnetic field. Following \cite{Willis2012}, this optimisation problem is written in dimensionless form as 
\begin{equation}
\begin{split}
    \max_{ \boldsymbol{B}_0,\boldsymbol{U} } \quad & \hat{J}_T(\boldsymbol{B}_0,\boldsymbol{U}) = \langle \boldsymbol{B}_T,\boldsymbol{B}_T \rangle \; \equiv \frac{1}{V}\int \boldsymbol{B}^2_T \; dV, \\
\hbox{s.t.} \quad & \langle \boldsymbol{B}_0,\boldsymbol{B}_0 \rangle = 1, \quad \langle \boldsymbol{U},\boldsymbol{U} \rangle = 1, \\
& \frac{\partial \boldsymbol{B} }{\partial t} - \nabla \times \big( \boldsymbol{U} \times \boldsymbol{B} \big) - \nabla \Pi -  Rm^{-1} \nabla^2 \boldsymbol{B} = 0, \\
& \nabla \cdot \boldsymbol{B} = 0, \quad \nabla \cdot \boldsymbol{U} = 0,
\end{split}
\label{eq:Max_kinematic}
\end{equation}
where $V$ is the volume of the box and $Rm$ (the magnetic Reynolds number) is a control parameter quantifying the relative strength of electromagnetic induction compared to ohmic dissipation. Due to the domain periodicity the problem can be spatially discretised using a Fourier basis. Spatial differentiation amounts to multiplication by $i {\bf k}$ in Fourier space (where $\bf k$ is the wavenumber) such that integration by parts discretely or continuously yield the same adjoint operators. As a result, only errors associated with the temporal discretization can arise in the gradient estimation.

Equation \eqref{eq:Max_kinematic} and its adjoint \eqref{eq:Euler_Lagrange_Kinematic_Adjoint} are time-integrated in the open-source code {\sc Dedalus} \cite{Burns2020} using a Crank-Nicolson Adams-Bashforth 1 (CNAB1) scheme. The problem setup and numerical resolutions are the same as in \cite{Willis2012} and the dimensionless control parameters are here both set to $Rm=T=1$. First we test the gradient with respect to the magnetic field $\delta \mathcal{L}/\delta \boldsymbol{B}_0$, holding $\boldsymbol{U}$ fixed as prescribed in \cite{Willis2012}, and subsequently the gradient with respect to the velocity field $\delta \mathcal{L}/\delta \boldsymbol{U}$ by holding $\boldsymbol{B}_0$ fixed. \autoref{table:Gradient_Test_UB} compares the gradient approximation obtained from the continuous adjoint equations \eqref{eq:Euler_Lagrange_Kinematic_Adjoint} with that of the discrete adjoint equations \eqref{eq:Disc_Kinematic_Euler_Lagrange_CNAB1} - \eqref{eq:Disc_Kinematic_Euler_Velocity}. Both are derived in terms of a Lagrangian formulation of \eqref{eq:Max_kinematic} in  \autoref{sec:Discrete_Adjoint}. To calculate the order of convergence when assessing the validity of \eqref{eq:Tay_R2_test}, we assume that for any $h_i$
\begin{equation}
\mathcal{W}(h_i) = | \hat{J}(\boldsymbol{X}_0 + h_i \delta \boldsymbol{X} ) - \hat{J}(\boldsymbol{X}_0) - h_i \langle \delta \boldsymbol{X}, \frac{\delta \mathcal{L}}{\delta \boldsymbol{X}_0} \rangle | \approxeq C |h_i|^{\gamma},
\label{eq:Order_determine}
\end{equation}
where $C$ and the exponent $\gamma$ are constants. Making the approximation \eqref{eq:Order_determine} for two different $h_i$ we estimate
\begin{equation}
\gamma \approxeq  \frac{ \log \big( \mathcal{W}(h_1)/ \mathcal{W}(h_2) \big) }{ \log(|h_1|/|h_2|) }.
\label{eq:Order_determine_exponent}
\end{equation} 
In all tables a significant improvement is seen when using the discrete adjoint while the continuous formulation cannot provide the mathematically consistent approximation demanded by \eqref{eq:Tay_R2_test}.

\begin{table}
\caption{\label{table:Gradient_Test_UB} Comparison of the continuous and discrete gradient for $N_{x,y,z} = 24$ Fourier modes, time-step $\Delta t =$1e-03 as per \cite{Willis2012}. (a) The velocity field $\boldsymbol{U}$ as given in \cite{Willis2012} is held fixed. (b) The initial magnetic field $\boldsymbol{B}_0$ is randomly chosen and held fixed. In both cases the vectors $\boldsymbol{B}_0, \delta \boldsymbol{B}, \delta \boldsymbol{U}$ of unit norm are generated by time-stepping vectors of noise a small number of iterations and then re-normalising. This ensures the boundary and divergence free conditions are satisfied whilst it avoids biasing the initial data.}
\begin{tabular}{ c | c || c | c || c | c }
\hline
(a)     &              & Continuous &          & Discrete $\mathcal{O}(\Delta t)$ & \\
\hline
$h$ & $|\hat{J}(\boldsymbol{B}_0 + h \delta \boldsymbol{B})-\hat{J}(\boldsymbol{B}_0)|/h$ & $| \langle \delta \boldsymbol{B}, \frac{\delta \mathcal{L}}{\delta \boldsymbol{B}_0} \rangle |$ & Order $\gamma$ & $| \langle \delta \boldsymbol{B}, \frac{\delta \mathcal{L}}{\delta \boldsymbol{B}_0} \rangle |$ & Order $\gamma$  \\
 \hline
 5.00e-05  & 5.60377e-03 & 5.60419e-03 & 0.41015 & 5.60363e-03 & 1.9999 \\
 2.50e-05  & 5.60370e-03 & 5.60419e-03 & 0.77621 & 5.60363e-03 & 2.0000 \\
 1.25e-05  & 5.60367e-03 & 5.60419e-03 & 0.89991 & 5.60363e-03 & 1.9999 \\
 6.25e-06  & \underline{5.6036}5e-03 & \underline{5.60}419e-03 & 0.95244 & \underline{5.6036}3e-03 & 2.0000 \\ 
 \hline
(b)        &              & Continuous &          & Discrete $\mathcal{O}(\Delta t)$ & \\
\hline
$h$ & $|\hat{J}(\boldsymbol{U} + h \delta \boldsymbol{U})-\hat{J}(\boldsymbol{U})|/h$ & $| \langle \delta \boldsymbol{U}, \frac{\delta \mathcal{L}}{\delta \boldsymbol{U}} \rangle |$ & Order $\gamma$ & $| \langle \delta \boldsymbol{U}, \frac{\delta \mathcal{L}}{\delta \boldsymbol{U}} \rangle |$ & Order $\gamma$  \\
 \hline
 5.00e-05  & 2.65775e-05 & 2.60762e-05 & 1.0024 & 2.65767e-05 & 2.0002 \\
 2.50e-05  & 2.65771e-05 & 2.60762e-05 & 1.0012 & 2.65767e-05 & 2.0009 \\ 
 1.25e-05  & 2.65769e-05 & 2.60762e-05 & 1.0006 & 2.65767e-05 & 2.0039  \\
 6.25e-06  & \underline{2.6576}8e-05 & \underline{2.6}0762e-05 & 1.0003  &  \underline{2.6576}7e-05 & 2.0226 \\ 
 \hline
\end{tabular}
\end{table}

\subsubsection{Second test-case: Swift-Hohenberg multi-stability}

The Swift-Hohenberg equation constitutes a paradigm for studying pattern formation in physics. It is known to have multiple stable solutions coexisting for the same parameter values, characterized by different energy levels \cite{Knobloch2015SpatialSystems}. This naturally raises the question of triggering transition from one stable state to another. Following \cite{Lecoanet_SHE} we consider the problem of selecting an initial condition $u_0 = u(x,t=0)$ of energy $E_0$, which when added to the stable, zero-energy equilibrium $u = 0$ induces a transition to a new, higher-energy equilibrium. This can be achieved by considering
\begin{equation}
\begin{split}
    \max_{ u_0 } \quad & \hat{J}(u) = \int_{t=0}^{t=T} \int_x |u(x,t)|^2 \; dx \; dt, \\
\hbox{s.t.} \quad & \int_x \frac{1}{2}|u_0|^2 \; dx = E_0, \\
\quad & \partial_t u + (1 + \partial_x^2)^2u -a u = 1.8u^2 - u^3,
\end{split}
\label{eq:SH23_Objective}
\end{equation}
on a periodic domain of dimensionless length $L = 12 \pi$ with parameter $a=-0.3$ and target time $T=50$ following \cite{Lecoanet_SHE}. These values ensure that the zero-energy state $u = 0$ is linearly stable to infinitesimal perturbations, but unstable to some finite-amplitude perturbations which can drive the system away to a different stable state. Discretising the spatial operators using a Fourier basis we once again ensure that only temporal errors can contaminate the gradient estimate. Equation \eqref{eq:SH23_Objective} and its discrete adjoint \eqref{eq:SH23_Disc_Euler_Lagrange_CNAB1} are time-integrated in {\sc Dedalus} using a first order backwards differencing (SBDF1) scheme.

\begin{table}[h!]
\caption{\label{table:Gradient_Test_SH} Test of gradient's accuracy for $N = 256, \Delta t = 0.05$ and an SBDF1 time-stepping scheme as per \cite{Lecoanet_SHE}. The discrete equations achieve a consistent gradient valid to 5 decimal places, however smaller $h$ will provide a yet more consistent approximation.}
\begin{tabular}{c | c || c | c || c | c }
\hline
         &              & Continuous &          & Discrete $\mathcal{O}(\Delta t)$ & \\
\hline
$h$ & $| \hat{J}( u + h \delta u ) - \hat{J}(u) |/h$ & $| \langle \delta u, \frac{\delta \mathcal{L}}{\delta u } \rangle |$ & Order $\gamma$ & $| \langle \delta u, \frac{\delta \mathcal{L}}{\delta u } \rangle |$ & Order $\gamma$  \\
 \hline
 5.000e-05  & 2.797111 & 3.171736 & 0.999860 & 2.797075 & 2.002582 \\
 2.500e-05  & 2.797093 & 3.171736 & 0.999930 & 2.797075 & 2.001365 \\ 
 1.250e-05  & 2.797084 & 3.171736 & 0.999965 & 2.797075 & 2.000622 \\
 6.125e-06  & \underline{2.7970}80 & 3.171736 & 0.999982 & \underline{2.7970}75 & 2.000059 \\
 \hline
\end{tabular}
\end{table}

\autoref{table:Gradient_Test_SH} reports a comparison of the continuous and discrete gradient by considering a fixed perturbations of noise for different $h$. While several digits of accuracy and a consistent gradient are obtained using the discrete equations we were unable to obtain a consistent gradient using the continuous equations for any choice of $\Delta t$ or $h$. Having demonstrated the superiority of the discrete adjoint approach on two spatially self-adjoint discretisations, we now consider a wall bounded problem in which both the temporal and spatial discretisation can contaminate the gradient.

\subsubsection{Third test-case: optimal mixing problem}\label{sec:Optimal_Mixing_Grad_Test}

In both the natural environment and industrial applications the mixing of fluids plays a crucial role. In the ocean for example, the irreversible mixing which occurs is the consequence of a complex interplay between density stratification and background shear playing out in a vast range of length-scales. A grand challenge in fluid dynamics community is to understand how the turbulent motions which ensue from these effects transport warm saline water, biomass and organisms such as phytoplankton across density surfaces in the oceans \cite{dauxois2021confronting}. An idealised example of optimal mixing is the optimal stirring strategy in a plane, pressure-driven channel flow \cite{foures2014optimal,marcotte2018optimal,heffernan2022robust}. This example considers what is, for example, the most efficient way of mixing a stably stratified layer of hot and cold fluid? This question may be posed as an optimisation problem, where the initial velocity perturbation $\boldsymbol{u}_0$ is determined so as to optimise well-mixedness. This can be achieved by minimising the following mix-norm for the density variation:
\begin{equation}
\begin{split}
\min_{ \boldsymbol{u}_0 } \quad & \hat{J}(\boldsymbol{u},b) = \frac{1}{2} \langle |\nabla^{-1} b(\boldsymbol{x},T)|^2 \rangle, \\
\hbox{s.t.} \quad & \frac{1}{2}\langle |\boldsymbol{u}_0|^2 \rangle = E_0, \\
\quad & \frac{\partial \boldsymbol{u}}{\partial t} + (\boldsymbol{U}_0 \cdot \nabla) \boldsymbol{u} + (\boldsymbol{u} \cdot \nabla)[\boldsymbol{U}_0 + \boldsymbol{u}] - \nabla p + Ri b \hat{\boldsymbol{z}} - \frac{1}{Re} \nabla^2 \boldsymbol{u}, \\
& \frac{\partial b}{\partial t} + ([\boldsymbol{U}_0 + \boldsymbol{u}] \cdot \nabla) b - \frac{1}{Pe} \nabla^2 b, \quad \nabla \cdot \boldsymbol{u} = 0, \\
\end{split}
\label{eq:Mixing_Objective}    
\end{equation}
where the non-dimensional Reynolds and P\'eclet numbers $Re,Pe = 500$ characterise the strength of flow inertia to viscous and thermal diffusion respectively and the Richardson number $Ri=0.05$ quantifies the strength of buoyancy relative to the background shear. Although we do not present it here, in \texttt{SphereManOpt} we include the additional test case where the time averaged kinetic energy is maximised instead, as in \cite{marcotte2018optimal}. We assume periodicity in the streamwise direction $x \in [0,4 \pi]$ while in the shearwise direction $z \in [-1,1]$ we impose no-slip velocity, the pressure gauge, the no-flux condition on the deviation density and its accompanying gauge
\begin{equation}
\boldsymbol{u}(x,z=\pm 1, t) = 0,  \; \int_V p \; dV = 0,  \quad \partial_z b(x,z=\pm 1, t) = 0,  \; \int_V b \; dV = 0.
\end{equation}
The latter gauge is essential as it (1) ensures that the mix-norm is well defined \cite{thiffeault2012using} and (2) prevents the optimisation from stalling due to an undefined gauge freedom. The background shear flow and the initial density deviation profile are given by
\begin{equation}
\boldsymbol{U}_0 = (1-z^2)\hat{\boldsymbol{x}}, \quad b(\boldsymbol{x},t=0) = -\frac{1}{2} \hbox{erf}(\frac{z}{\delta_0}), \quad \hbox{with} \quad \delta_0 = 0.125,
\end{equation}
a value larger than that of previous studies to facilitate the use of a spectral method. Following \cite{marcotte2018optimal} the initial amplitude is set to $E_0 = 0.02$ and the optimisation window to $T=5$. This slightly shorter optimisation window is chosen so as to avoid excessively small values of the mix-norm which were found to induce rounding errors. \eqref{eq:Mixing_Objective} and its discrete adjoint \eqref{eq:Euler_Lagrange_Mixing_Adjoint} are time-integrated in {\sc Dedalus} using a first order backwards differencing (SBDF1). In contrast to the previous examples the spatial operators are discretised using a Fourier basis to treat the streamwise direction and a Chebyshev basis in the shearwise direction. Consequently the discrete operators appearing in \eqref{eq:Disc_Euler_Lagrange_CNAB1_0}-\eqref{eq:Disc_Euler_Lagrange_CNAB1} are no longer self-adjoint, such that both temporal and spatial errors can now contaminate the gradient estimate.

\begin{table}[h]
\caption{\label{table:Gradient_Test_Mixing} Test of gradient's accuracy for $N_x,N_z = 256,128, \Delta t = 1e-03$ using an SBDF1 scheme. The initial perturbation velocities $\boldsymbol{u},\delta \boldsymbol{u}$ of unit norm are generated by time-stepping vectors of noise a small number of iterations and then re-normalising.}
\begin{tabular}{c | c || c | c || c | c }
\hline
         &              & Continuous &          & Discrete $\mathcal{O}(\Delta t)$ & \\
\hline
$h$ & $|\hat{J}( \boldsymbol{u} + h \delta \boldsymbol{u})-\hat{J}(\boldsymbol{u})|/h$ & $| \langle \delta \boldsymbol{u}, \frac{\delta \mathcal{L}}{\delta \boldsymbol{u} } \rangle |$ & Order $\gamma$ & $| \langle \delta \boldsymbol{u}, \frac{\delta \mathcal{L}}{\delta \boldsymbol{u} } \rangle |$ & Order $\gamma$  \\
 \hline
 \hline
 5.00e-05  & 7.51875e-03 & 1.4081e-02 & 1.0079 & 7.51848e-03 & 1.9999 \\
 2.50e-05  & 7.51862e-03 & 1.4081e-02 & 1.0040 & 7.51848e-03 & 1.9995 \\
 1.25e-05  & 7.51855e-03 & 1.4081e-02 & 1.0020 & 7.51848e-03 & 1.9974 \\
     6.25e-06  & \underline{7.518}52e-03 & 1.4081e-02 & 1.0010 & \underline{7.518}48e-03 & 1.9886 \\
 \hline
 \hline
\end{tabular}
\end{table}

\autoref{table:Gradient_Test_Mixing} reports a comparison of the continuous and discrete gradient by considering a fixed perturbations of noise for different $h$ and for different objective functions. We find that several digits of accuracy and a consistent gradient are obtained using the discrete equations while we were unable to obtain a consistent gradient using the continuous equations for any choice of $\Delta t$ or $h$. 

\subsection{Run time}\label{sec:Test_Timing}

\begin{table}
\caption{ \label{table:Gradient_Timings} Ratio of the adjoint integration time versus the forwards integration time as computed using a single node with 2 AMD Epyc 7302 @ 3 GHz – 16 cores processors. To minimise memory usage only the forwards solve in spectral space is stored.}
\begin{tabular}{c | c | c | c }
\hline
Problem     & Forward Solve (cpu-s) & Adjoint Solve (cpu-s) & Ratio \\
\hline
 Kinematic dynamo   & 50.432  & 91.94  & 1.823 \\
 Swift-Hohenberg    & 2.473   & 0.388   & 0.156 \\ 
 Optimal mixing     & 2794.185 & 2476.635 & 0.886 \\
\hline
\end{tabular}
\end{table}

\autoref{table:Gradient_Timings} reports the efficiency of this section's gradient tests. The adjoint of the kinematic dynamo problem having twice the forward solve's dimension costs almost double as anticipated. The optimal mixing and Swift-Hohenberg problems cost equal or substantially less. Computations require the forwards solve in coefficient space only, thus minimising memory usage as compared with automatic differentiation.


\section{Gradient descent on a spherical manifold}\label{sec:Linesearch_Algorithim}

In the previous section we detailed how to obtain a numerically consistent approximation of the gradient of the pure functional $\hat{J}(\boldsymbol{X}_0)$. Extending this further we now consider
\begin{equation}
    \min_{\hat{\boldsymbol{X}}_0 \in \mathcal{S}} \quad \hat{J}(\hat{\boldsymbol{X}}_0) = J(\boldsymbol{X}(\hat{\boldsymbol{X}}_0),\hat{\boldsymbol{X}}_0),
    \label{eq:constrained_opt_unit_norm} 
\end{equation}
where
\begin{equation}
\mathcal{S} = \{ \hat{\boldsymbol{X}} \; | \; || \hat{\boldsymbol{X}} ||^2 = E_0 \},
\label{eq:Manifold_definition}    
\end{equation}
such that, by restricting $\boldsymbol{\hat{X}}_0$ to the spherical manifold $\mathcal{S}$, the equality constraints are built-in. By defining \eqref{eq:constrained_opt_unit_norm} as an unconstrained optimisation problem we can now proceed in outlining an appropriate gradient descent procedure, either by using the rotation method \cite{Douglas1998} (also termed exponential map in the literature on optimisation on Riemannian manifolds \cite{absil2009optimization})
which moves $\hat{\boldsymbol{X}}_0$ along geodesics of $\mathcal{S}$ in the search direction $\boldsymbol{d}_k$, or using the retraction formulae of \cite{absil2009optimization} 
which is computationally simpler and easily generalises to the consideration of $L^p$ norms \cite{sato2022riemannian} (our discussion concentrates exclusively on $L^2$-norms).

In the fluid mechanics community, the rotation method was first used to perform norm-constrained optimisation on a nonlinear PDE problem by \cite{foures2013localization} who combined it with a Polak-Ribi\`ere conjugate gradient method \cite{hager2006survey}. They achieved faster convergence with the rotation method than with the Lagrange multiplier method. More recently \cite{skene_yeh_schmid_taira_2022} highlighted this link and utilised Riemannian optimisation techniques for performing a unit-norm constrained linear optimisation problem. In what follows we outline the recent theoretical developments concerning steepest descent \cite{boumal2020introduction} and conjugate-gradient descent developed by \cite{sato2021riemannian_book,sato2021riemannian_article}. This is followed by our numerical validation of their formulae, which using a stringent line-search is shown to provide convergence consistent with theoretical guarantees.

\subsection{Conjugate gradient formulation}\label{sec:SD_and_CG_descent}

Following \cite{boumal2020introduction,sato2021riemannian_book} we denote the parameter vector at iterate $k$ by $\hat{\boldsymbol{X}}_k$, the Euclidean gradient of the objective function by $\nabla \hat{J}(\hat{\boldsymbol{X}}_k)$ and the tangent-gradient of the objective function (i.e. its gradient on the spherical manifold $\mathcal{S}$) by
\begin{equation}
    \boldsymbol{g}_k = \bigg(I - \frac{ \boldsymbol{\hat{X}}_k \boldsymbol{\hat{X}}_k^T }{|| \boldsymbol{\hat{X}}_k||^2} \bigg)\nabla \hat{J}(\hat{\boldsymbol{X}}_k).
    \label{eq:tangent_gradient}
\end{equation}
Starting from an initial guess $\hat{\boldsymbol{X}}_0$, restricted to $\mathcal{S}$, gradient descent proceeds via the recurrence
\begin{equation}
\begin{split}
\boldsymbol{\hat{X}}_{k+1} & = \frac{ \big( \boldsymbol{\hat{X}}_k + \alpha_k \boldsymbol{d}_k \big) }{|| \boldsymbol{\hat{X}}_k + \alpha_k \boldsymbol{d}_k ||} \sqrt{E_0}, \\ %
\boldsymbol{d}_k           & = -\boldsymbol{g}_k + \beta_k \; s_k \; \mathcal{T}_{\boldsymbol{\hat{X}}_k} \big( \boldsymbol{d}_{k-1} \big),    
\end{split}
\label{eq:Update_procedure}    
\end{equation}
where $\beta_k$ is the conjugate-gradient update parameter \cite{sato2021riemannian_article,sato2021riemannian_book} and the step size $\alpha_k$ is determined using a line-search procedure in order to find an improved minimum of $\hat{J}(\hat{\boldsymbol{X}}_0)$. For steepest-descent $\beta_k = 0$ is set, while in the conjugate-gradient method we use a combined Riemannian Polak-Ribi\`ere and Fletcher-Reeves type update following a general formulation put forward by \cite{sato2021riemannian_article}, who demonstrated that the following choice of $\beta_k$ yields global convergence of the conjugate-gradient algorithm: 
\begin{equation}
    \beta_k = \hbox{max}\{ 0, \hbox{min}\{ \beta^{R-PR}_k, \beta^{R-FR}_k \} \},
    \label{eq:beta_parameter}
\end{equation}
with
\begin{equation}
\beta^{R-PR}_k = \frac{ \langle \boldsymbol{g}_k, \boldsymbol{g}_k -  \mathcal{T}_{\boldsymbol{\hat{X}}_k}( \boldsymbol{g}_{k-1} ) \rangle }{ || \boldsymbol{g}_{k-1} ||^2 }, \quad
\beta^{R-FR}_k = \frac{ || \boldsymbol{g}_k ||^2 }{ || \boldsymbol{g}_{k-1} ||^2 }.
\label{eq:Fletcher-Reeves_Polar-Ribere}
\end{equation}
The vector transport operator 
\begin{equation}
\mathcal{T}_{\boldsymbol{\hat{X}}_k}(\boldsymbol{d}_{k-1}) = \frac{\sqrt{E_0}}{|| \boldsymbol{\hat{X}}_k ||} \bigg( I - \frac{\boldsymbol{\hat{X}}_k \boldsymbol{\hat{X}}_k^T}{ || \boldsymbol{\hat{X}}_k||^2 } \bigg) \boldsymbol{d}_{k-1}, 
\end{equation}
and scaling coefficient
\begin{equation}
s_k =
\begin{cases}
\frac{||\boldsymbol{d}_{k-1} ||}{||\mathcal{T}_{\boldsymbol{\hat{X}}_k}(\boldsymbol{d}_{k-1})||}, \quad \hbox{if}  \quad ||\mathcal{T}_{\boldsymbol{\hat{X}}_k}(\boldsymbol{d}_{k-1})|| >  ||\boldsymbol{d}_{k-1} ||, \\
1, \quad\quad\quad\quad\quad \; \; \; \hbox{if}  \quad ||\mathcal{T}_{\boldsymbol{\hat{X}}_k}(\boldsymbol{d}_{k-1})|| \leq ||\boldsymbol{d}_{k-1} ||,
\end{cases}
\label{eq:transport_inequality}
\end{equation}
are used to account for the following: (1) vector addition on the manifold $\mathcal{S}$ only makes sense if the vectors live in the same tangent space, and (2) the inequality \eqref{eq:transport_inequality} must be satisfied for convergence of \eqref{eq:Update_procedure}; that is the transport operator must not increase the vectors norm \cite{sato2021riemannian_article}. In all cases the iterative procedure repeats until the gradient residual
\begin{equation}
    r = || \boldsymbol{g}_k|| = || \nabla \hat{J}_k || \sin(\theta_k), \quad \hbox{where} \quad \cos(\theta_k) = \frac{ \langle \boldsymbol{\hat{X}}_k, \nabla \hat{J}_k \rangle }{ || \boldsymbol{\hat{X}}_k || \; || \nabla \hat{J}_k || },
    \label{eq:Grad_Res_Error}
\end{equation}
involving the measure of the angle between the parameter vector $\boldsymbol{\hat{X}}_k$ and Euclidean gradient $\nabla \hat{J}_k$, becomes less than some numerically prescribed tolerance, meaning that no further progress can be made without departing too much from the constraint manifold. We note that this measure provides an immediate connection to, both the Lagrangian method \eqref{eq:necessary_condition} and to the Armijo condition \eqref{eq:Armijo} when $\boldsymbol{d}_k \approx -\boldsymbol{g}_k$. 

\subsection{Conditions on the step size}\label{step size}

Convergence guarantees on the scheme \eqref{eq:Update_procedure} can be achieved by imposing conditions on the step-size $\alpha$ \cite{sato2021riemannian_article,sato2021riemannian_book}. Assuming that $\hat{J}(\hat{\boldsymbol{X}}_K)$ is continuously differentiable, bounded from below and that the gradient is Lipschitz continuous for all $\boldsymbol{\hat{X}}_k \in \mathcal{S}$, the (Riemannian) strong Wolfe-conditions
\begin{eqnarray}
\hat{J}(\hat{\boldsymbol{X}}_{k+1}(\alpha_k)) \leq \hat{J}(\hat{\boldsymbol{X}}_k) + c_1 \alpha_k \langle \boldsymbol{g}_k,\boldsymbol{d}_k \rangle, \label{eq:Armijo} \\
 | \langle \boldsymbol{g}_{k+1}(\alpha_k), \mathcal{T}_{\hat{\boldsymbol{X}}_{k+1}} ( \boldsymbol{d}_k ) \rangle | \leq c_2 | \langle \boldsymbol{g}_k,\boldsymbol{d}_k \rangle |, \label{eq:Wolfe_Conditions_I_II}
\end{eqnarray}
with $0 < c_1 < c_2 < 1/2$ ensure global convergence to a stationary point \cite{sato2021riemannian_article}. Here we have used the notation $\boldsymbol{\hat{X}}_{k+1}(\alpha_k)$ to denote the step-size dependent choice of parameter vector and $\boldsymbol{g}_{k+1}(\alpha_k)$ to denote its corresponding tangent-gradient. In contrast to Euclidean gradient descent, the Riemannian Wolfe-conditions include an additional transport of the search direction $\mathcal{T}_{\boldsymbol{\hat{X}}_{k+1}}(\boldsymbol{d}_{k})$ to ensure that the inner product, which is now a metric on $\mathcal{S}$, remains well defined.

\subsection{Principle component analysis}

\begin{figure}[h]
    \centering
    \begin{subfigure}[b]{0.495\textwidth}
     \centering
    \includegraphics[scale=0.35]{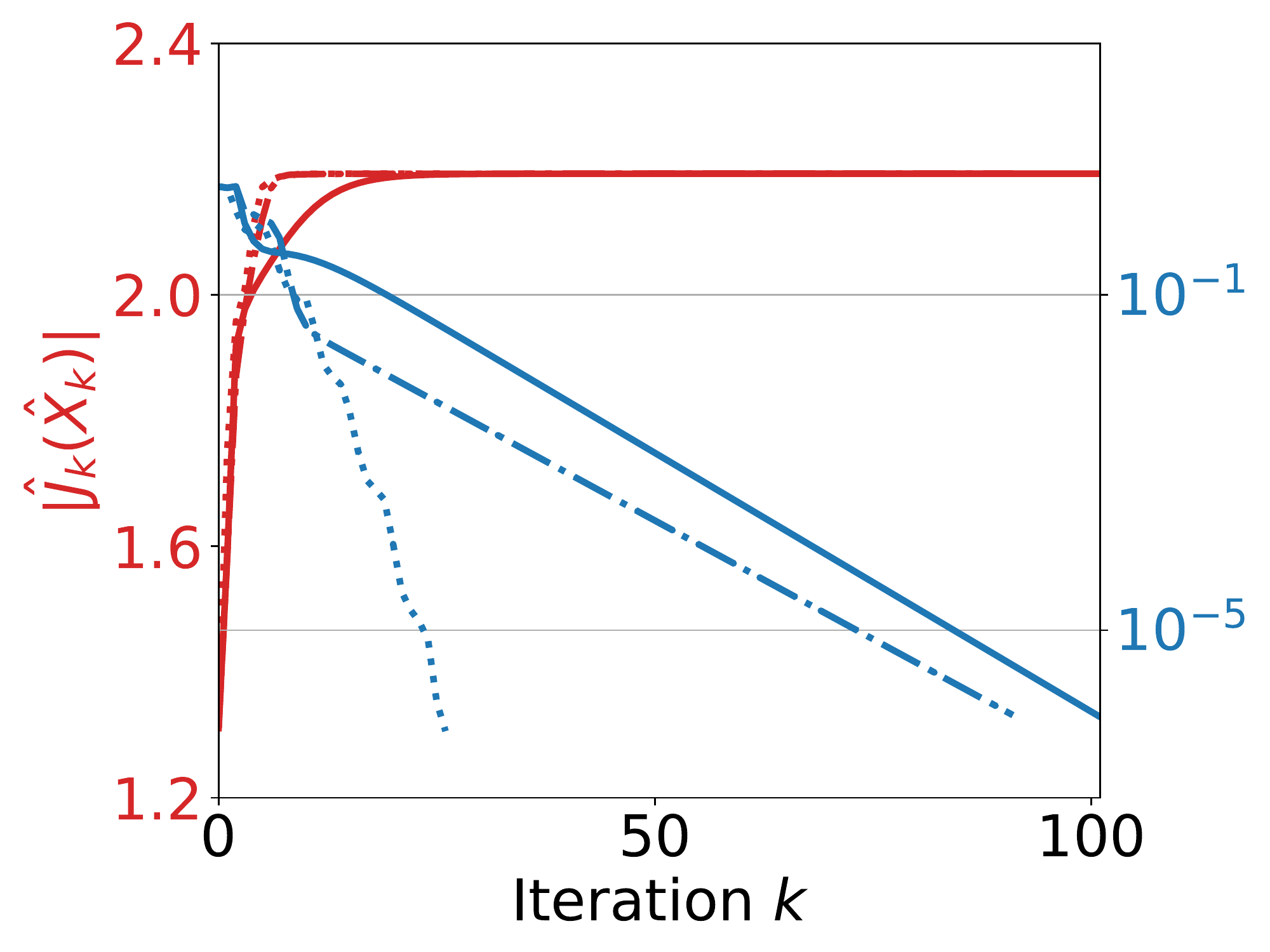}
    \caption{$N = 10$}
    \end{subfigure}
    \begin{subfigure}[b]{0.495\textwidth}
     \centering
     \includegraphics[scale=0.35]{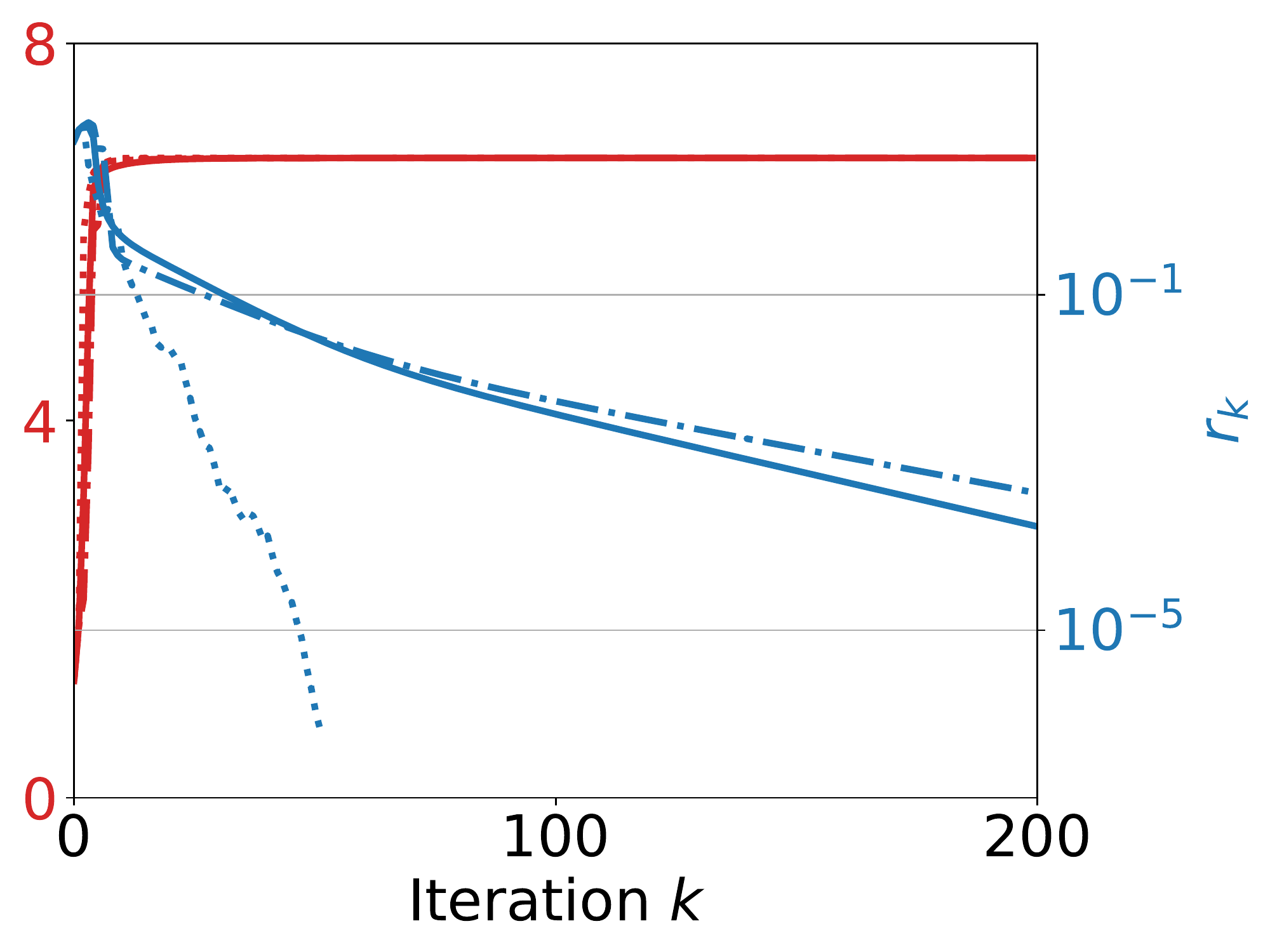}
     \caption{$N = 100$}
    \end{subfigure}
     \caption{Convergence history of the optimisation \eqref{eq:PCA_Objective} using steepest-descent with an Armijo line-search (solid line), conjugate-gradient with an Armijo line-search (chained line) and conjugate-gradient with a Wolfe I \& II line-search (dotted line) method. The absolute value of the objective function $|\hat{J}_k|$ is shown in red and the gradient residual $r_k$ in blue. Despite conducting a more expensive line-search the conjugate-gradient method, which exhibits super-linear convergence, ultimately requires fewer iterations to converge (see \autoref{table:Optimisation_Test_PCA_Results}).}
     \label{fig:PCA_CONV_Angle}
\end{figure}

To assess the efficacy of the update procedure and the practical cost of enforcing \eqref{eq:Armijo},\eqref{eq:Wolfe_Conditions_I_II}, we solve
\begin{equation}
\begin{split}
\min_{\boldsymbol{X}} \quad & \hat{J}(\boldsymbol{X})  = -\frac{1}{2} \boldsymbol{X}^T M \boldsymbol{X}, \\
\hbox{s.t.} \quad & ||\boldsymbol{X}||^2  = 1, \quad \hbox{where} \quad M = M^T, \quad \boldsymbol{X}^T M \boldsymbol{X} > 0,    
\end{split}
\label{eq:PCA_Objective}    
\end{equation}
which is equivalent to principle component analysis (PCA) or if $M$ corresponds to the discretisation of a self-adjoint linear operator, finding its leading eigenvector. Assuming an exact line-search, the rate of convergence of gradient descent to a stationary point follows analytically from the condition number $\kappa(M) = \lambda_n/\lambda_1$ of the Hessian where $\lambda_1,\lambda_n$ refers to the smallest/largest eigenvalue of $M$ respectively. Denoting the solution of \eqref{eq:PCA_Objective} by $\boldsymbol{X}^*$ the result of \cite{luenberger1973introduction} states for steepest descent
\begin{equation}
\hat{J}(\hat{\boldsymbol{X}}_{k+1}) - \hat{J}(\boldsymbol{X}^*) \leq  r^2 \big[ \hat{J}(\hat{\boldsymbol{X}}_k) - \hat{J}(\boldsymbol{X}^*) \big], \quad \hbox{where} \quad r \in \bigg] \frac{\kappa - 1}{\kappa + 1}, 1 \bigg[,
\label{eq:descent_bound}
\end{equation}
for large $k$, such that $\hat{\boldsymbol{X}}_k \to \boldsymbol{X}^*$ at a linear rate. To test this bound we generate random $M \in \mathbb{R}^{N \times N}$ and perform gradient descent. For $N=100$ we calculate $(\kappa(M)-1/\kappa(M)+1)^2 = 0.9799$ and using the slope of $(\hat{J}_{k+1} - \hat{J}^*)/(\hat{J}_k - \hat{J}^*) = 0.9187$, showing that the calculated estimate obeys the bound \eqref{eq:descent_bound} and monotonic linear convergence is observed. Performing the same optimisation using the conjugate gradient method \eqref{eq:Update_procedure} subject to the strong Wolfe conditions we obtained super-linear convergence as shown in \autoref{table:Optimisation_Test_PCA_Results} and \autoref{fig:PCA_CONV_Angle}.

\begin{table}[h]
\caption{\label{table:Optimisation_Test_PCA_Results} Comparison of the steepest descent and conjugate gradient methods applied to \eqref{eq:PCA_Objective}. We set the max step-size $\alpha_{max}=1$ and terminate optimisation when $r \leq$ 1e-06 or when iterations $>200$. Star $*$ denotes a stalled run. Despite conducting a more expensive line-search the conjugate-gradient routine converges super-linearly thus requiring fewer evaluations as shown in \autoref{fig:PCA_CONV_Angle}.}
\begin{tabular}{ c | c|  c | c| c | c }
 \hline
Routine & Size $N$ & line-search &  Obj evals $\hat{J}$ & Grad evals $\nabla \hat{J}$ & Residual $r$  \\
 \hline
 Gradient descent   & 10 & Armijo  & 105 & 102 & 9.2117e-07 \\
 \hline
 Conjugate gradient & 10 & Armijo  & 95 & 92 & 9.6906e-07 \\
 \hline
 Conjugate gradient & 10 & Wolfe I \& II & 51 & 34 & 6.2045e-07 \\ 
 \hline
 Gradient descent   & 100 & Armijo  & 205 & 200  & *2.9087e-05  \\
 \hline
 Conjugate gradient & 100 & Armijo  & 205 & 200 & *1.5407e-04 \\
 \hline
 Conjugate gradient & 100 & Wolfe I \& II  & 94 & 66  & 7.4820e-07 \\
 \hline
\end{tabular}
\end{table}

To facilitate the application of the descent routine outlined in this section, we have combined the update formula of \autoref{sec:SD_and_CG_descent} with an implementation of an Armijo \eqref{eq:Armijo} line-search based on cubic interpolation and an implementation of a strong Wolfe condition line-search  \eqref{eq:Wolfe_Conditions_I_II} following \cite{more1994line,wright1999numerical}. This is available in the companion package to this paper \texttt{SphereManOpt}  \cite{SphereManOpt2022} which also contains all examples presented in this paper. In contrast to the libraries \texttt{manopt} and \texttt{pymanopt} \cite{manopt,JMLR:v17:16-177}, our library focuses exclusively on spherical manifolds of variable radii (as desirable in fluid dynamics applications) and implements a line-search routine satisfying the strong Wolfe conditions. Such a line-search while necessary to ensure global convergence of the conjugate gradient method \cite{sato2021riemannian_book} is currently unavailable in \texttt{manopt} or \texttt{pymanopt} \cite{manopt,JMLR:v17:16-177}.


\section{Numerical performance}\label{sec:Numerical_Comparisons}

Having outlined the calculation of a consistent gradient estimate in \autoref{sec:Discrete_Adjoint} and a line-search following \cite{sato2021riemannian_book,sato2021riemannian_article} with global convergence to a stationary point in \autoref{sec:Linesearch_Algorithim}, we now combine the developments of these sections and apply them to solving for the optimal solution of the examples previously considered in \autoref{sec:Discrete_Adjoint}.

\subsection{Swift-Hohenberg multi-stability}\label{sec:SH23_performance}

For the choice of parameters given in \autoref{sec:Discrete_Adjoint} the Swift-Hohenberg equation has four stable equilibria. Following \cite{Lecoanet_SHE}, we find the smallest perturbation $u_0$ of energy $M_2$ which, when added to the zero-energy state $O$, triggers a transition to the next least-energy state $u_T$ of energy $S_2$. (In practice, $M_2$ is determined by repeatedly computing the optimal $u_0$, of gradually decreasing energy $M$, that maximises the energy growth by target time $T$. Since larger energies characterize the searched-for transition, $M_2$ corresponds to the smallest value of $M$ for which the optimal $u_0$ still triggers the transition.)

\begin{figure}[ht]
    \centering
    \includegraphics[scale=0.4]{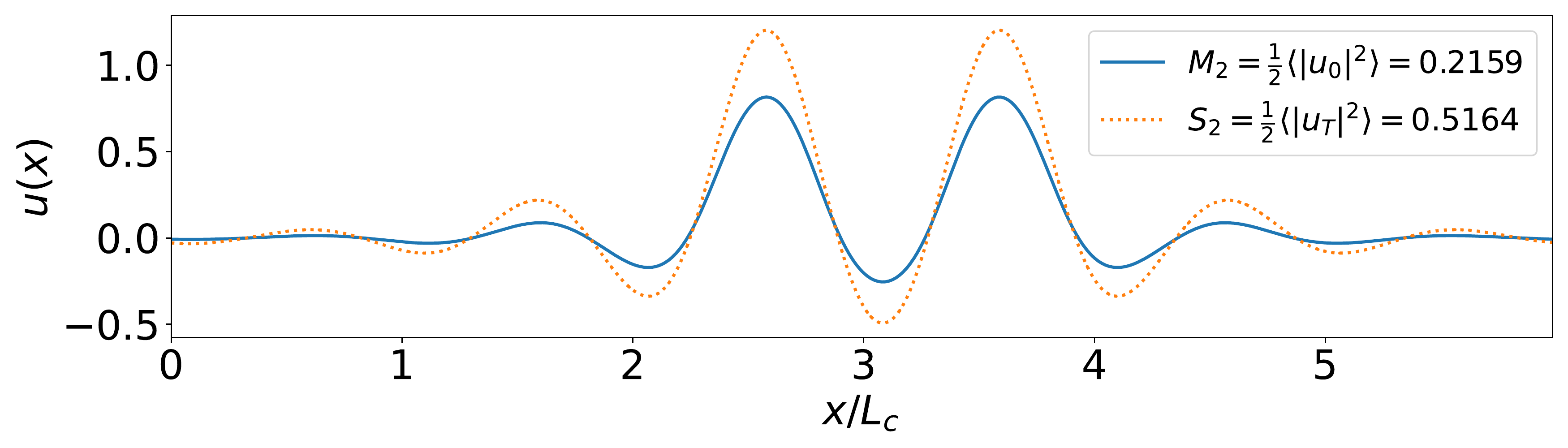}
    \vspace{-5mm}
    \caption{Optimisation using the discrete gradient recovers the energies of the initial condition $M_2$ and equilibrium state $S_2$ (inset). These are given in figures 1 and 5 of \cite{Lecoanet_SHE} as $M_2,S_2=(0.2048,0.5164)$.}
     \label{fig:DAL_CONV_SH23_Spatial_Result}
\end{figure}

\autoref{fig:DAL_CONV_SH23_Spatial_Result} compares amplitudes of the perturbation $M_2$ and stable equilibrium $S_2$ with the results of \cite{Lecoanet_SHE}. All steady equilibrium are correctly identified to four digits of precision while the perturbation amplitudes are correct to one digit. The differences with the result of \cite{Lecoanet_SHE} are explained by the fact we performed a small number of runs initiated from random noise, to probe a cost functional with numerous local minima. In contrast they perform several hundred runs so as to explore many possible basins of attraction.

\subsubsection{Discrete vs. continuous adjoint}

Having established that our code faithfully reproduces the results of \cite{Lecoanet_SHE} we now compare the performance of the discrete and continuous gradient when optimising for the perturbation $u_0$. The convergence of these runs are reported in rows (a),(c) of \autoref{table:Optimisation_Test_SH23}. Both runs use steepest-descent (SD) with an Armijo line-search and are initiated using random noise. While the continuous gradient seems to converge faster to a small residual, this appearance is in fact misleading. By running the optimum (found with the `continuous', i.e. differentiate-then-discretize approach) once through the direct and discrete adjoint solvers (i.e. discretize-then-differentiate), we obtain the corresponding `discrete residual' which is in fact much higher. This is highlighted in row (a) of \autoref{table:Optimisation_Test_SH23} where the discrete residual is 6 orders of magnitude greater that the continuous one, showing that only the discrete gradient can accurately yield convergence to a stationary point. A further disadvantage of the continuous gradient as shown in row (b) of  \autoref{table:Optimisation_Test_SH23} is that when supplied to the conjugate-gradient (CG) routine it stalls, thus eliminating the possibility of super-linear convergence.  

\subsubsection{Line-search algorithms}

\begin{figure}[ht]
    \centering
    \begin{subfigure}{0.325\textwidth}
     \centering
     \includegraphics[scale=0.28]{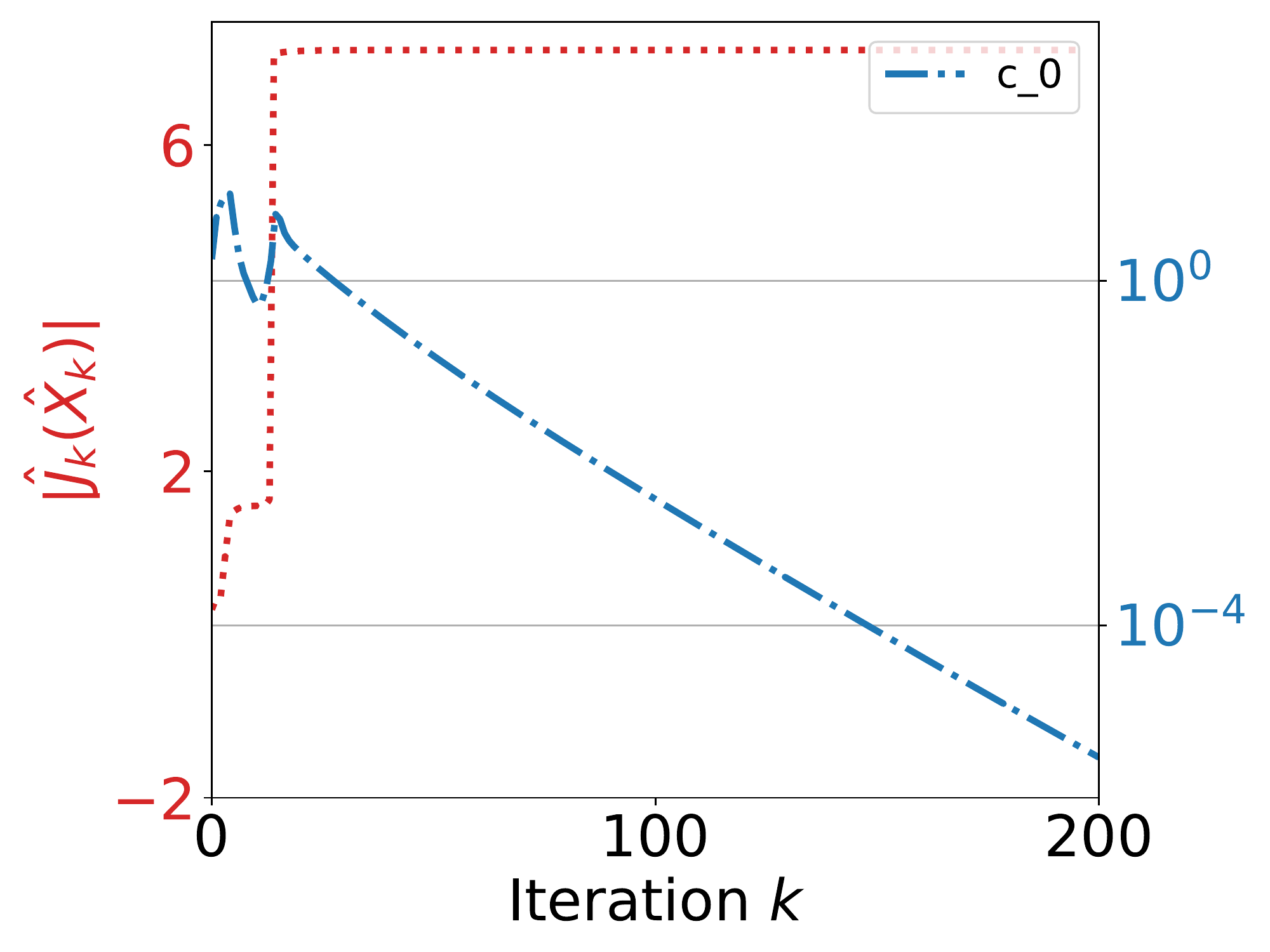}
     \caption*{(d) SD (Armijo)}
    \end{subfigure}
    \begin{subfigure}{0.325\textwidth}
     \centering
     \includegraphics[scale=0.28]{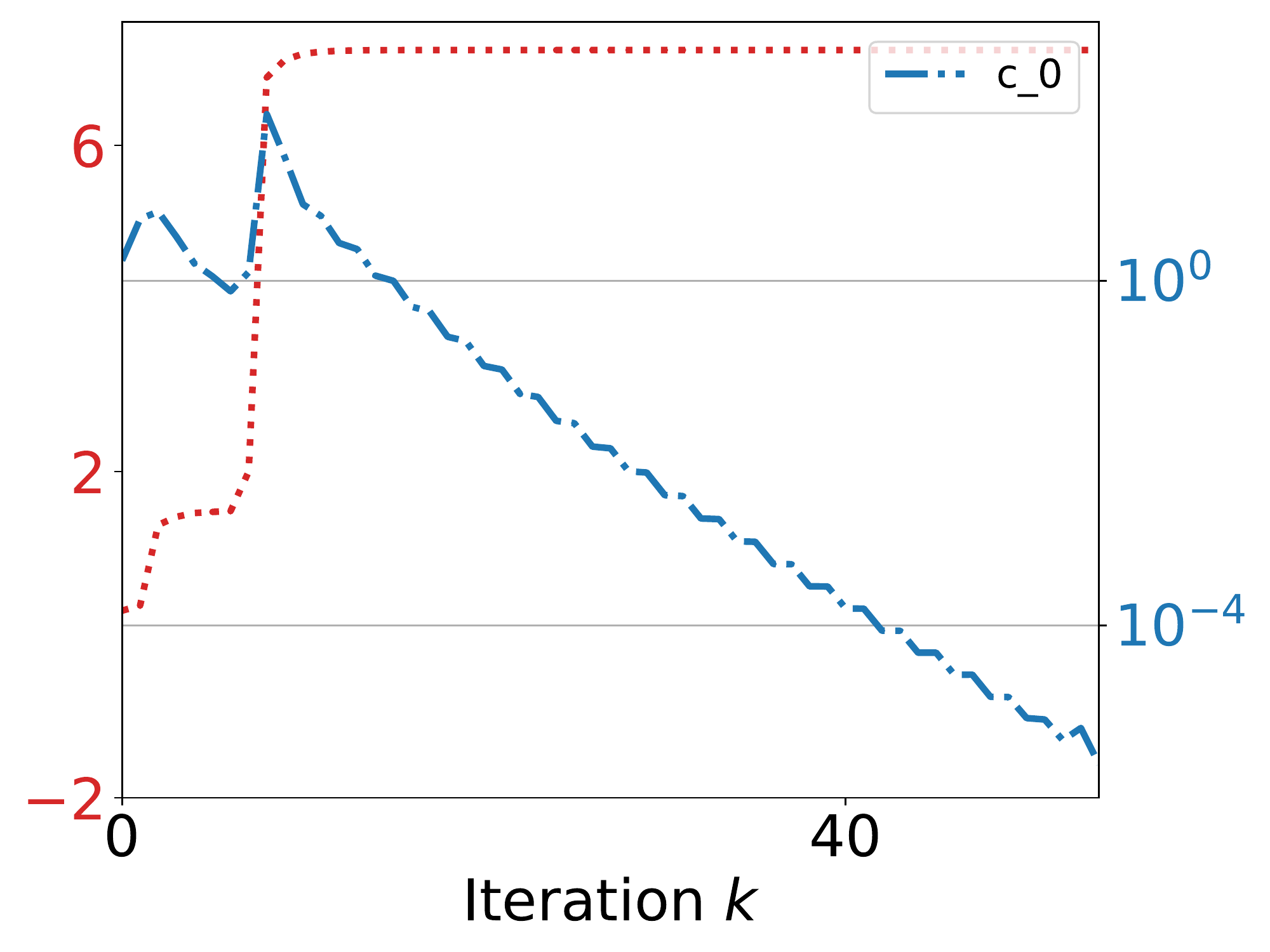}
     \caption*{(e) SD (Wolfe I \& II)}
    \end{subfigure}
    \begin{subfigure}{0.325\textwidth}
     \centering
     \includegraphics[scale=0.28]{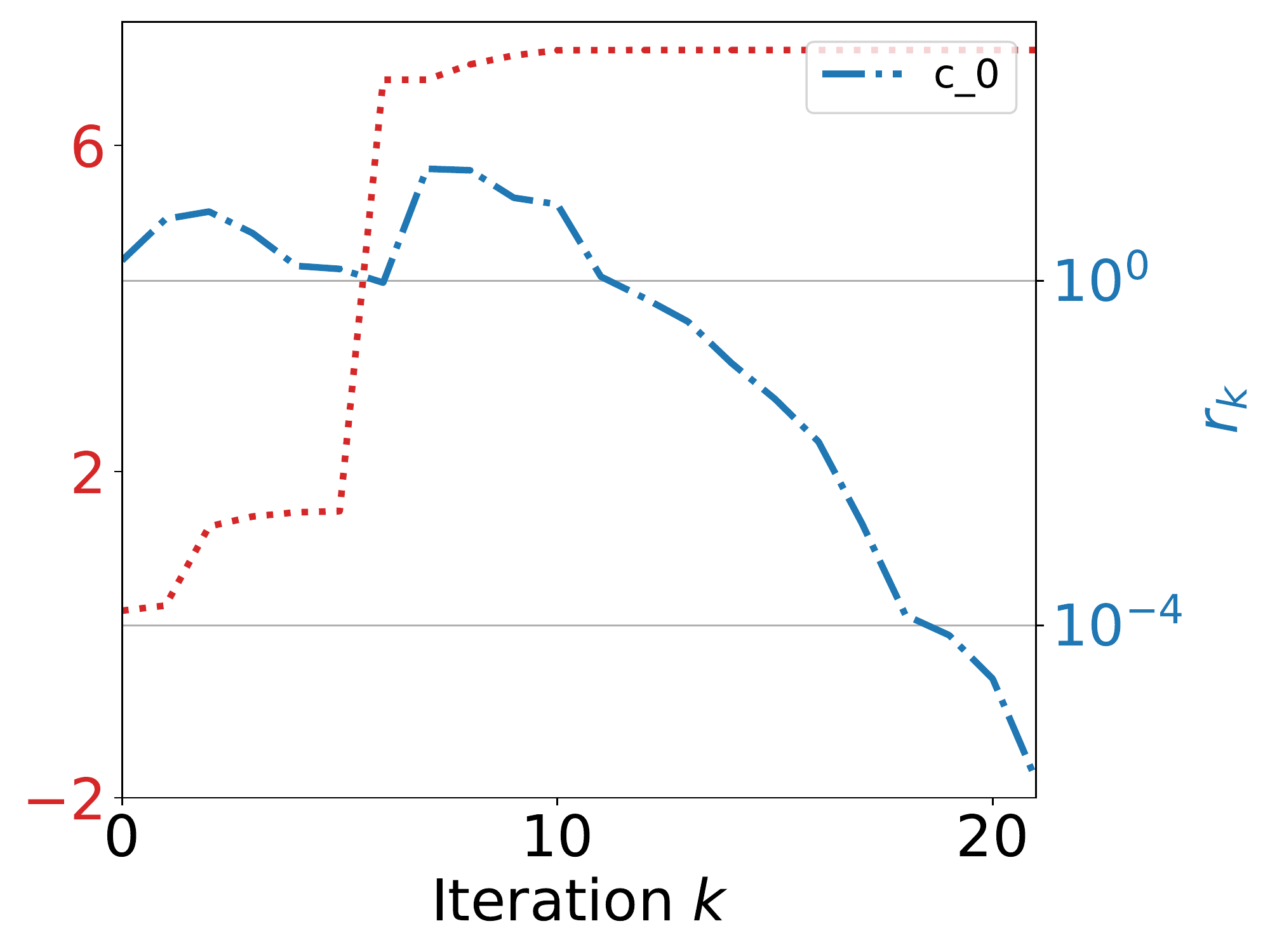}
     \caption*{(g) CG (Wolfe I \& II)}
    \end{subfigure}
     \caption{Convergence of the optimisation runs in rows (d)-(g) of \autoref{table:Optimisation_Test_SH23}, in terms of the cost functional $J$ (dotted, red) and gradient residual $r$ (chained, blue), to the perturbation $M_2$ when initiated with random noise of the same amplitude. The steepest descent (SD) methods (d),(e) converge linearly while the conjugate-gradient (CG) method (g) super-linearly, thus demanding fewer iterations.}
     \label{fig:DAL_CONV_SH_Angle}
\end{figure}

With the superiority of the discrete gradient established we now consider the efficiency and convergence behaviour of three different descent routines. The convergence of each routine is reported in rows (c)-(f) of  \autoref{table:Optimisation_Test_SH23} and frames (c)-(f) of \autoref{fig:DAL_CONV_SH_Angle}. While all routines reach a small residual error, the most commonly used routine, a backtracking line-search accompanied by an Armijo condition, performs the worst. As anticipated the conjugate-gradient method implementing the strong Wolfe conditions and thus enjoying super-linear convergence performs best.

\begin{table}[ht]
\caption{\label{table:Optimisation_Test_SH23} Test summary and numerical cost of reaching a stationary point using different descent routines as shown in \autoref{fig:DAL_CONV_SH_Angle}. We set the max step-size $\alpha_{max}=\pi$ and terminate optimisation when $r \leq$ 1e-06 or when iterations $>200$. Star $*$ denotes a stalled run. In rows (a)-(c) we have reported the \texttt{continuous residual} as well as the discrete residual. }
\begin{tabular}{ c | c|  c | c| c | c }
 \hline
Routine & Gradient & line-search &  Obj evals $\hat{J}(\hat{\boldsymbol{u}}_0)$ & Grad evals $\frac{\delta \mathcal{L}}{\delta \boldsymbol{u}_0}$ & Residual  $r$ \\
 \hline
 (a) SD   & Continuous & Armijo  & 163 & 173  & 1.5190/$\mathtt{9.2551e-07}$ \\
  \hline
  (b) CG & Continuous & Wolfe I \& II & 63 & 37   & *1.4662/$\mathtt{1.1718e-01}$ \\
 \hline
 (c) CG & Continuous & Armijo & 217 & 200   & *3.7429/$\mathtt{2.8598}$ \\
 \hline
 (d) SD   & Discrete   & Armijo  & 213 & 197  & 2.9400e-06  \\
 \hline
 (e) SD   & Discrete   & Wolfe I \& II  & 118 & 92  & 2.4488e-06  \\
 \hline
 (f) CG & Discrete   & Armijo & 29 & 13  & *3.8577e+01 \\
 \hline
 (g) CG & Discrete   & Wolfe I \& II & 91 & 44  & 1.6003e-06 \\
 \hline
\end{tabular}
\end{table}

\subsection{Kinematic dynamo}

 \begin{figure}[h!]
    \centering
    \begin{subfigure}[b]{0.495\textwidth}
     \centering
    \includegraphics[width=0.85\textwidth]{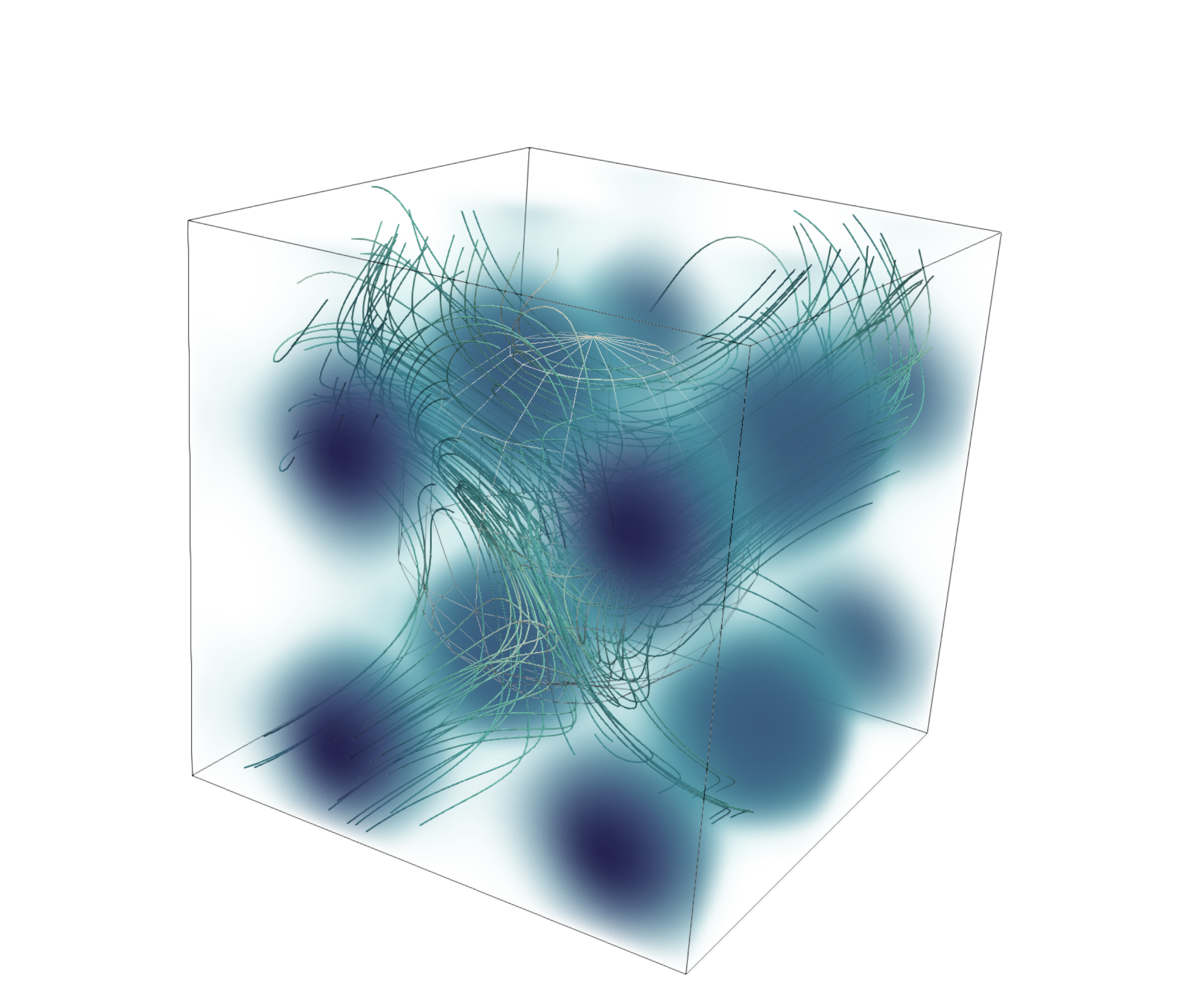}
    \caption{Magnetic field $\boldsymbol{B}_T$}
    \end{subfigure}
    \hfill
    \begin{subfigure}[b]{0.495\textwidth}
     \centering
     \includegraphics[width=0.85\textwidth]{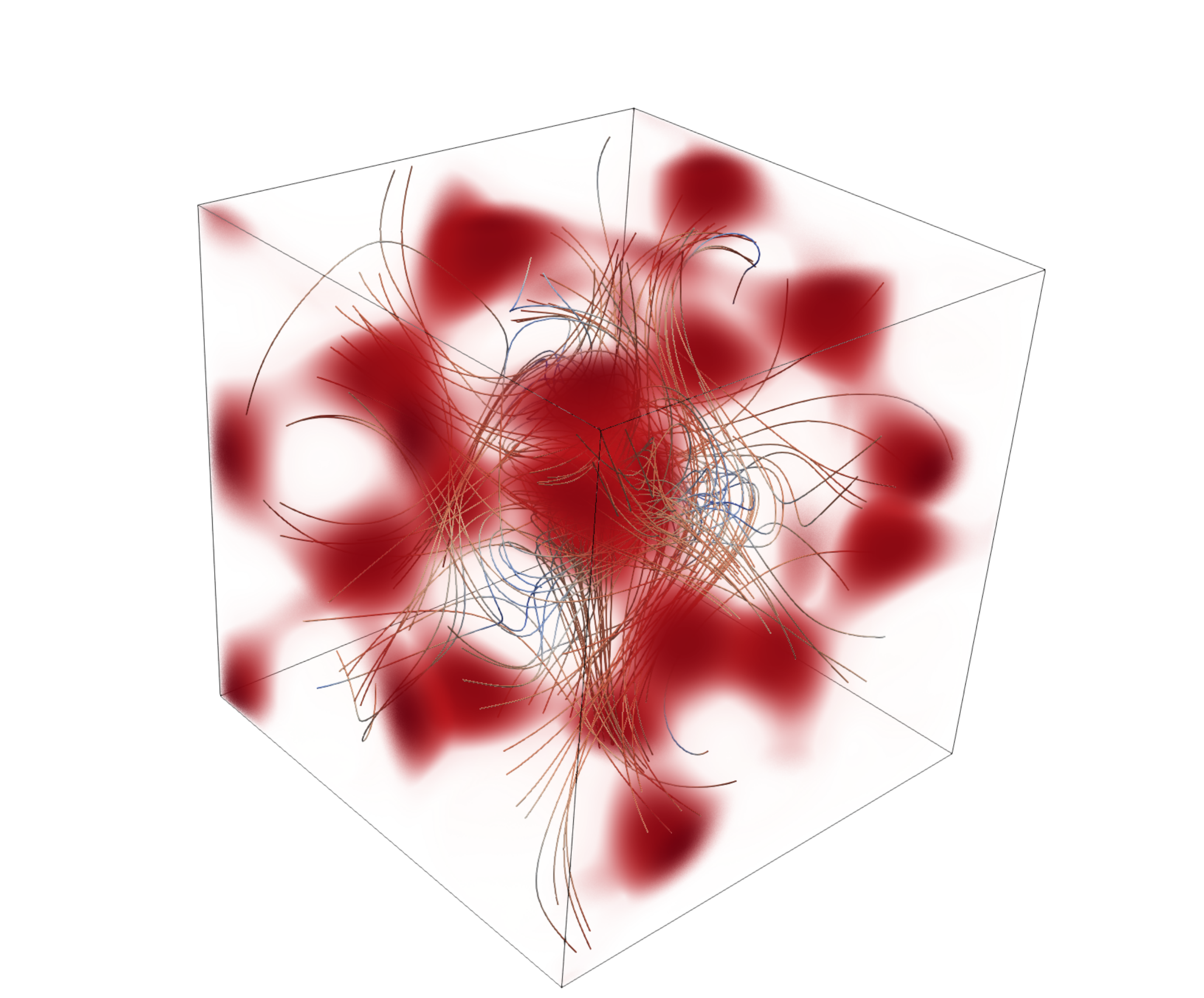}
     \caption{Velocity field $\boldsymbol{U}$}
    \end{subfigure}
     \caption{The optimal (a) magnetic and (b) velocity field for $Rm=T=1$ which yield $\langle \boldsymbol{B}_T^2 \rangle = 0.4329$. To be compared with figures 1 and 2 of \cite{Willis2012}. This optimum was calculated using an Armijo based line-search corresponding to row (c) \autoref{table:Optimisation_Test_UB}.}
     \label{fig:DAL_CONV_Spatial_Result}
\end{figure}

Having established the validity of the gradient in \ref{sec:Test_cases}, where only the discrete adjoint gives a consistent estimate, and similarly the validity of our code by comparison with \cite{Willis2012}(see \autoref{fig:DAL_CONV_Spatial_Result}), we now first demonstrate the superiority of the discrete gradient when optimising for fields $\boldsymbol{U}$ and $\boldsymbol{B}_0$. In contrast to the previous examples this demands choosing a single step-size $\alpha_k$ to simultaneously update all $n$ components of the parameter vector $\hat{\boldsymbol{X}}_k$, whilst retaining a termination condition based on the residual $r^i_k$ of each component. We argue that this is preferable to taking a summed quantity such as $r_k = \sum_{i=0}^n r^i_k$ which could lead to over and under converged directions compensating one-another. One weakness of the outlined approach is that when different components of $\hat{\boldsymbol{X}}_k$ converge at different rates, we will limit the convergence rate by restricting ourselves to a single step-size. Preconditioning the update step presents a possible solution to this problem \cite{hager2006survey}.

\subsubsection{Discrete vs. continuous adjoint}

\begin{figure}[h!]
    \centering
    \begin{subfigure}[b]{0.495\textwidth}
     \centering
    \includegraphics[width=0.85\textwidth]{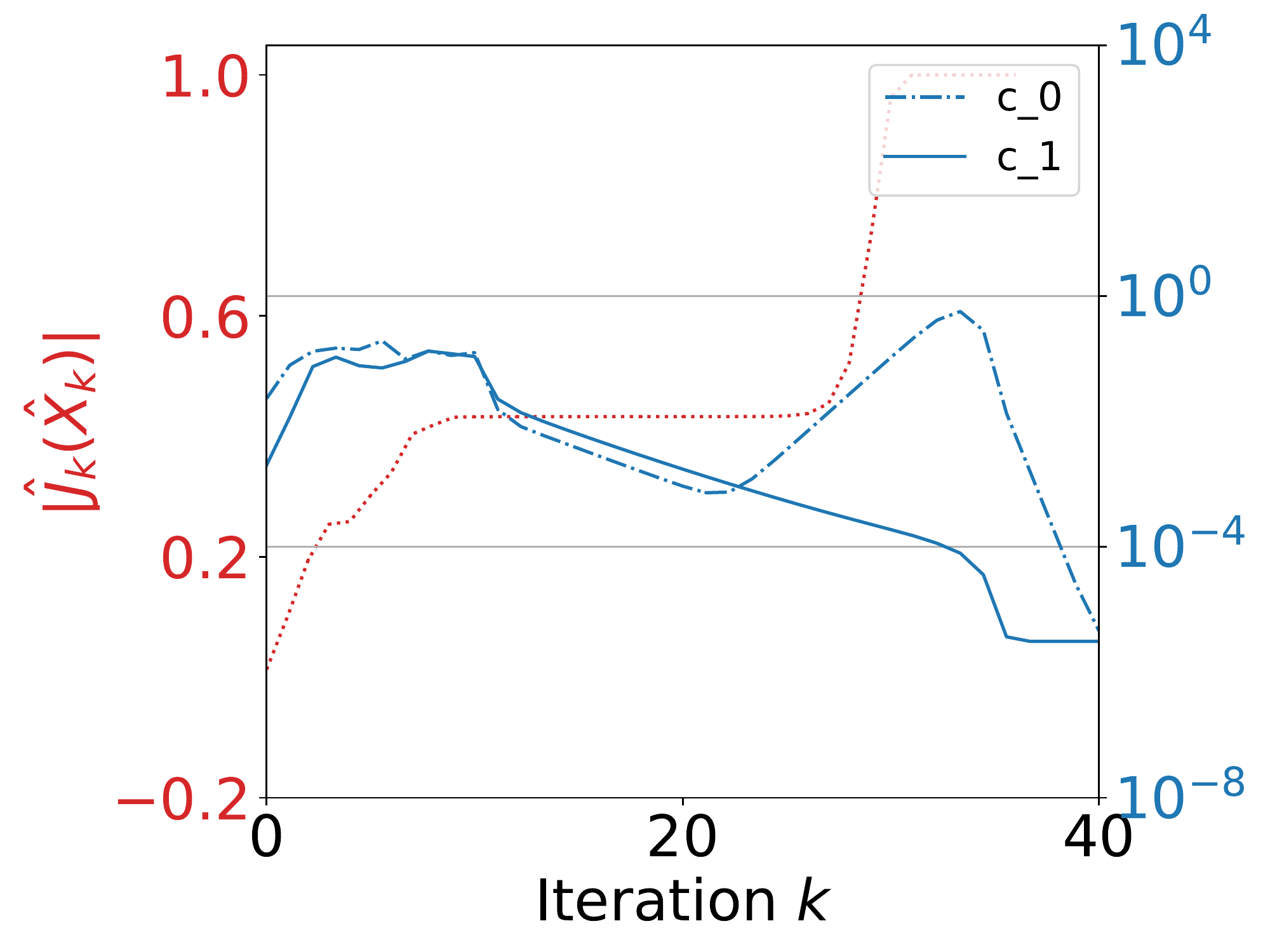}
     \caption*{(a) SD (Armijo)}
    \end{subfigure}
    \hfill
    \begin{subfigure}[b]{0.495\textwidth}
     \centering
     \includegraphics[width=0.85\textwidth]{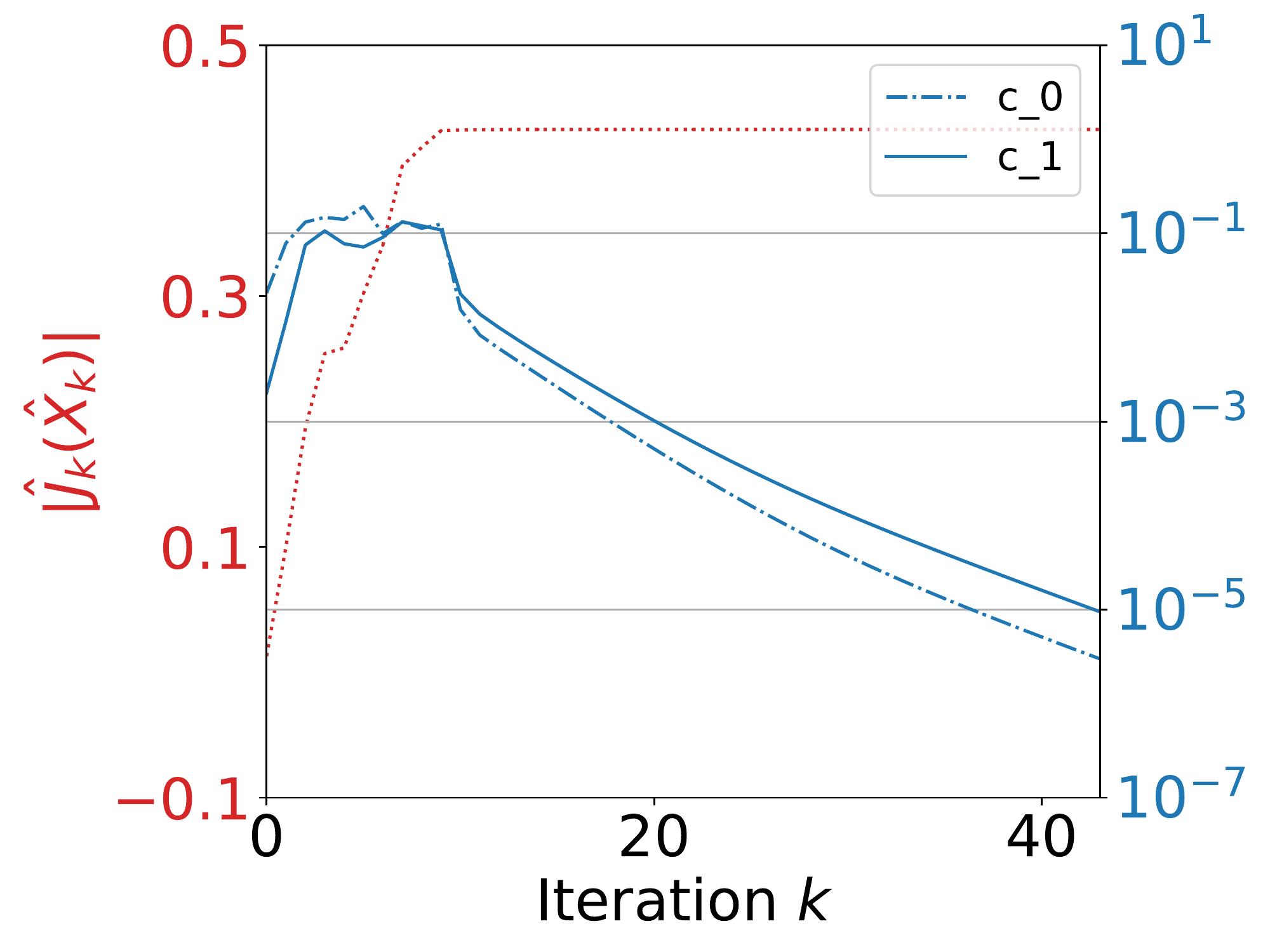}
     \caption*{(c) SD (Armijo)}
    \end{subfigure}
    
    \begin{subfigure}[b]{0.495\textwidth}
     \centering
     \includegraphics[width=0.85\textwidth]{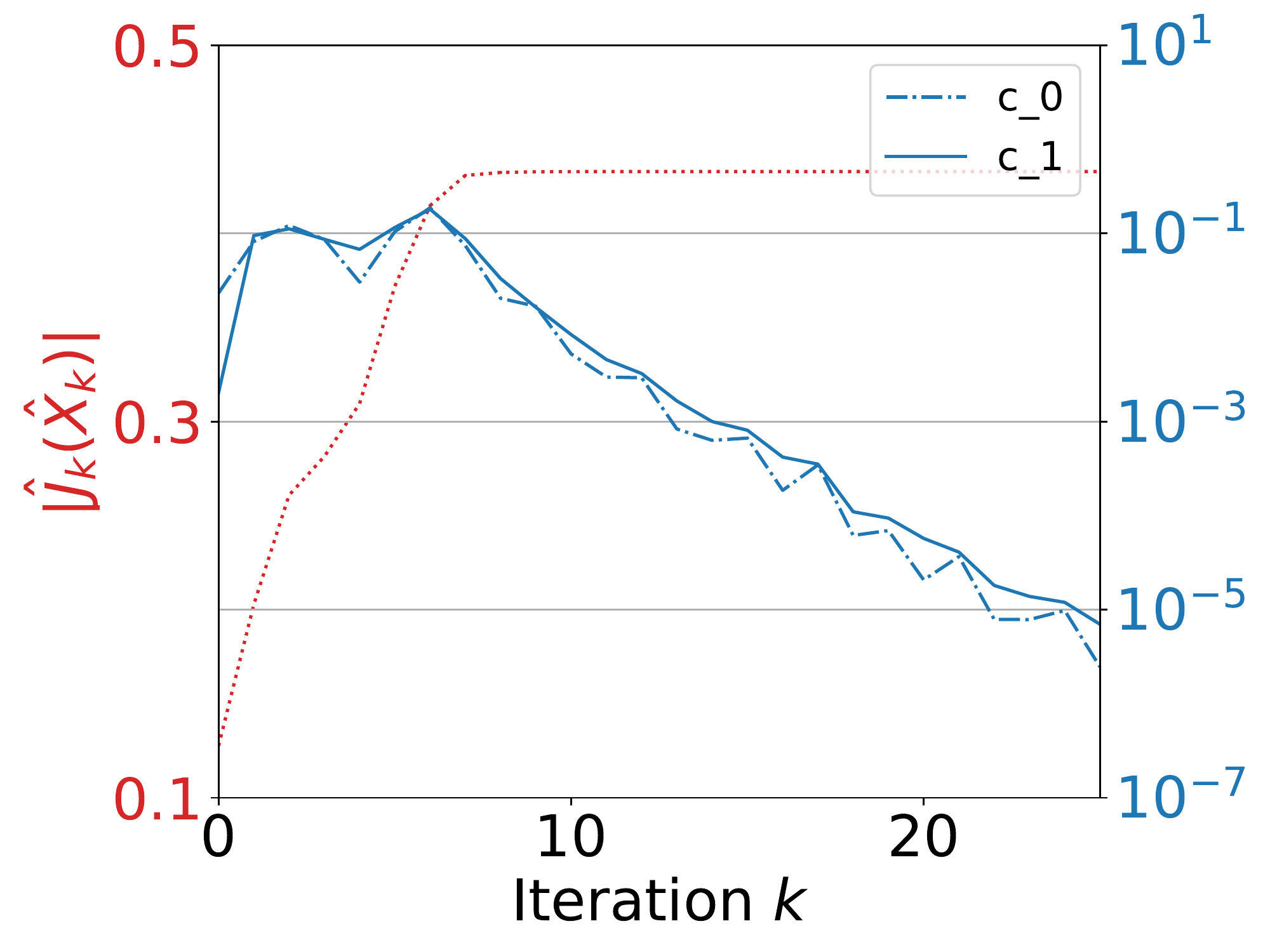}
     \caption*{(d) SD (Wolfe I \& II)}
    \end{subfigure}
    \begin{subfigure}[b]{0.495\textwidth}
     \centering
     \includegraphics[width=0.85\textwidth]{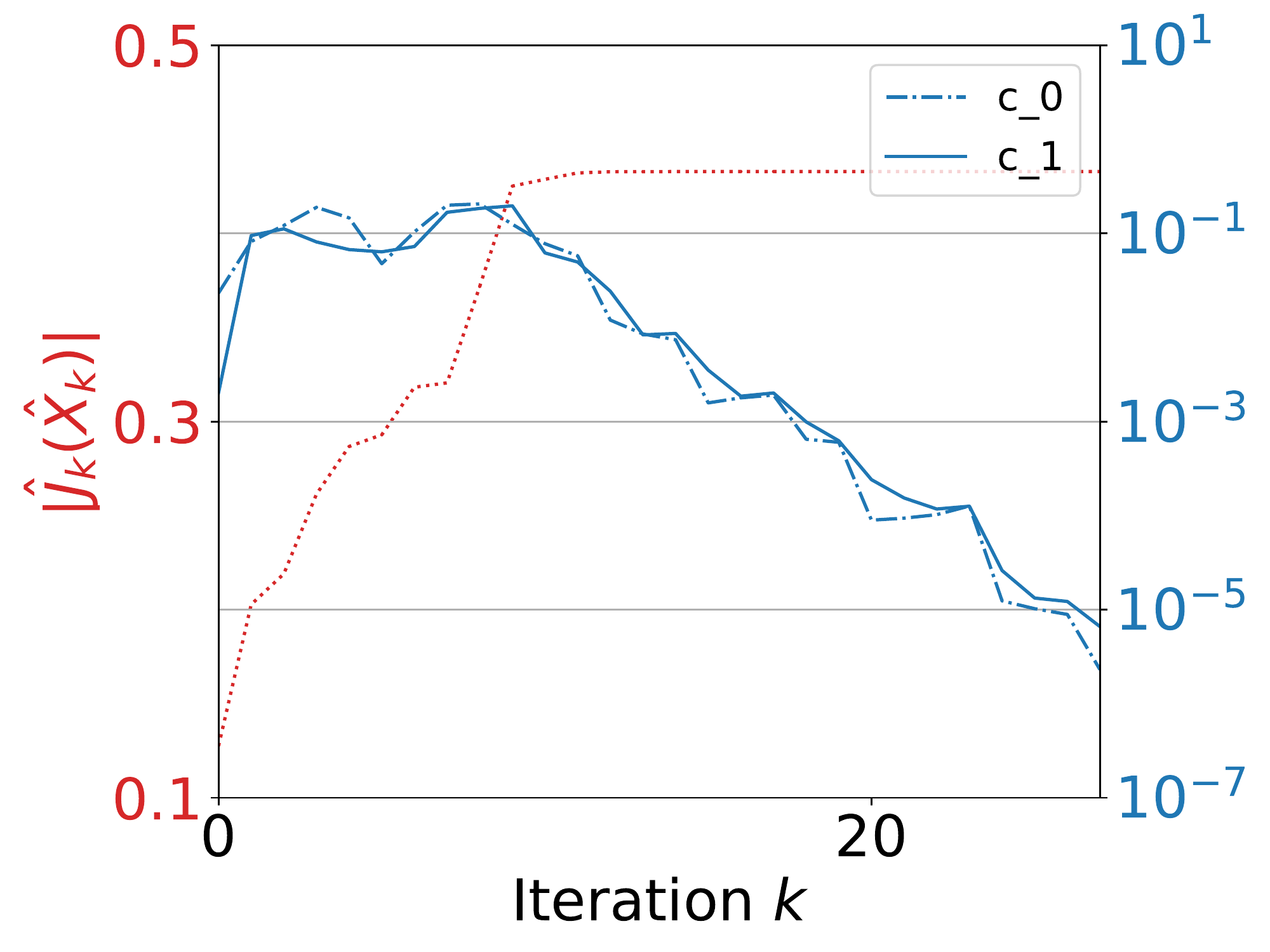}
     \caption*{(f) CG (Wolfe I \& II)}
    \end{subfigure}

     \caption{Convergence history in terms of the residual errors $r^B_k$ (shown as $c_0$) and $r^U_k$ (shown as $c_1$). Optimisation (a) used a continuous gradient and (c-f) a discrete gradient while their routines and convergence cost are summarised in \autoref{table:Optimisation_Test_UB}. While the continuous gradient appears to converge faster it is in fact to a spurious solution with non-zero divergence and net magnetic flux. Routines implementing the Wolfe conditions converge in fewer iterations but their line-search procedures are slightly more costly.}
     \label{fig:DAL_CONV_KDYN}
\end{figure}

Rows (a) and (c) of \autoref{table:Optimisation_Test_UB} compare the performance of the continuous and discrete gradient using steepest descent (SD) combined with an Armijo line-search. While the continuous gradient appears to converge in fewer gradient/function evaluations, we find that upon diverging after $\sim 25$ iterations it does so to a spurious solution characterised by a non-zero magnetic divergence and net magnetic flux (\autoref{fig:DAL_CONV_KDYN}(a)). The optimisation using the discrete gradient by contrast converges monotonically to the desired convergence criteria \autoref{fig:DAL_CONV_KDYN}(c). As previous it was not possible to avail of the conjugate-gradient (CG) method when using a continuous gradient (row (b) \autoref{table:Optimisation_Test_UB}).

\begin{table}[ht]
\caption{\label{table:Optimisation_Test_UB} Numerical cost of reaching a stationary point using different descent routines as shown in \autoref{fig:DAL_CONV_KDYN}. We set $\alpha_{max}=100$ and terminate optimisation
when $r^i_k \leq$ 1e-06 or when iterations $>200$. Star $*$ denotes a stalled run, while $**$ a non-physical solution. All rows are given in terms of their discrete gradient residual.}
\begin{tabular}{ c | c|  c | c| c | c | c }
 \hline
Routine & Gradient & Line-search & evals $\hat{J}(\hat{\boldsymbol{B}}_0,\hat{\boldsymbol{U}})$ & evals $\frac{\delta \mathcal{L}}{\delta \boldsymbol{U}},\frac{\delta \mathcal{L}}{\delta \boldsymbol{B}_0}$ & $r^B_k$ & $r^U_k$   \\
\hline
(a) SD & Continuous & Armijo               & 42  & 37 & **1.2519e-05 & **3.0875e-06 \\
\hline
(b) CG & Continuous & Wolfe I \& II                & 2  & 7 & *2.3210e-02 & *2.0199e-03 \\
\hline
(c) SD & Discrete & Armijo & 48 & 44 & 2.9980e-06 & 9.4994e-06 \\
\hline
(d) SD & Discrete & Wolfe I \& II & 53 & 53 & 2.4662e-06 & 6.9963e-06 \\
\hline
(e) CG & Discrete & Armijo & 55  & 42 & 3.1154e-06 &  9.7692e-06 \\
\hline
(f) CG & Discrete & Wolfe I \& II & 61  & 57 & 2.3031e-06 &  6.6113e-06 \\
\hline
\end{tabular}
\end{table}

\subsubsection{Line-search algorithms}

Having established the necessity of the discrete gradient, we now consider the efficiency of the strong Wolfe line-search and conjugate-gradient routines when using a combined update approach. Columns (d) and (f) of \autoref{table:Optimisation_Test_UB} repeat the same calculation using a strong Wolfe line-search with and without the conjugate-gradient update. While both routines converge in fewer iterations when comparing \autoref{fig:DAL_CONV_KDYN}(c) with figures \ref{fig:DAL_CONV_KDYN}(d,f) their total number of function and gradient evaluations is however slightly larger.

\subsection{Optimal Mixing}

The previous examples demonstrate that our gradient descent algorithm efficiently achieves a high degree of convergence, but that this is possible only when using the discrete gradient. We now combine all this paper's developments to treat a PDE discretised using a combined Fourier/Chebyshev basis, as is required to solve the optimal mixing problem \eqref{eq:Mixing_Objective}. As of yet adjoint optimisation studies with a discrete gradient have used finite difference \cite{vishnampet2015practical}, finite element \cite{Zahr2016,Mitusch2019} or finite volume methods \cite{schmid2017stability} and in doing so encountered some difficulties in treating boundary conditions and pressure gauge conditions. By using the {\sc Dedalus} code which treats the pressure and divergence free condition (and similar integral conditions) alongside all other variables during a single time-step no splitting methods or complications arise in our implementation \eqref{App:Preconditioned_First_Order_Scheme}. Given the prevalence of flow features with large gradients such as velocity boundary layers or density fronts, the accurate solution of control problems using a Chebyshev pseudospectral method is an essential step.

 \begin{figure}[h!]
    \centering
    \begin{subfigure}[b]{0.495\textwidth}
     \centering
    \includegraphics[width=0.9\textwidth]{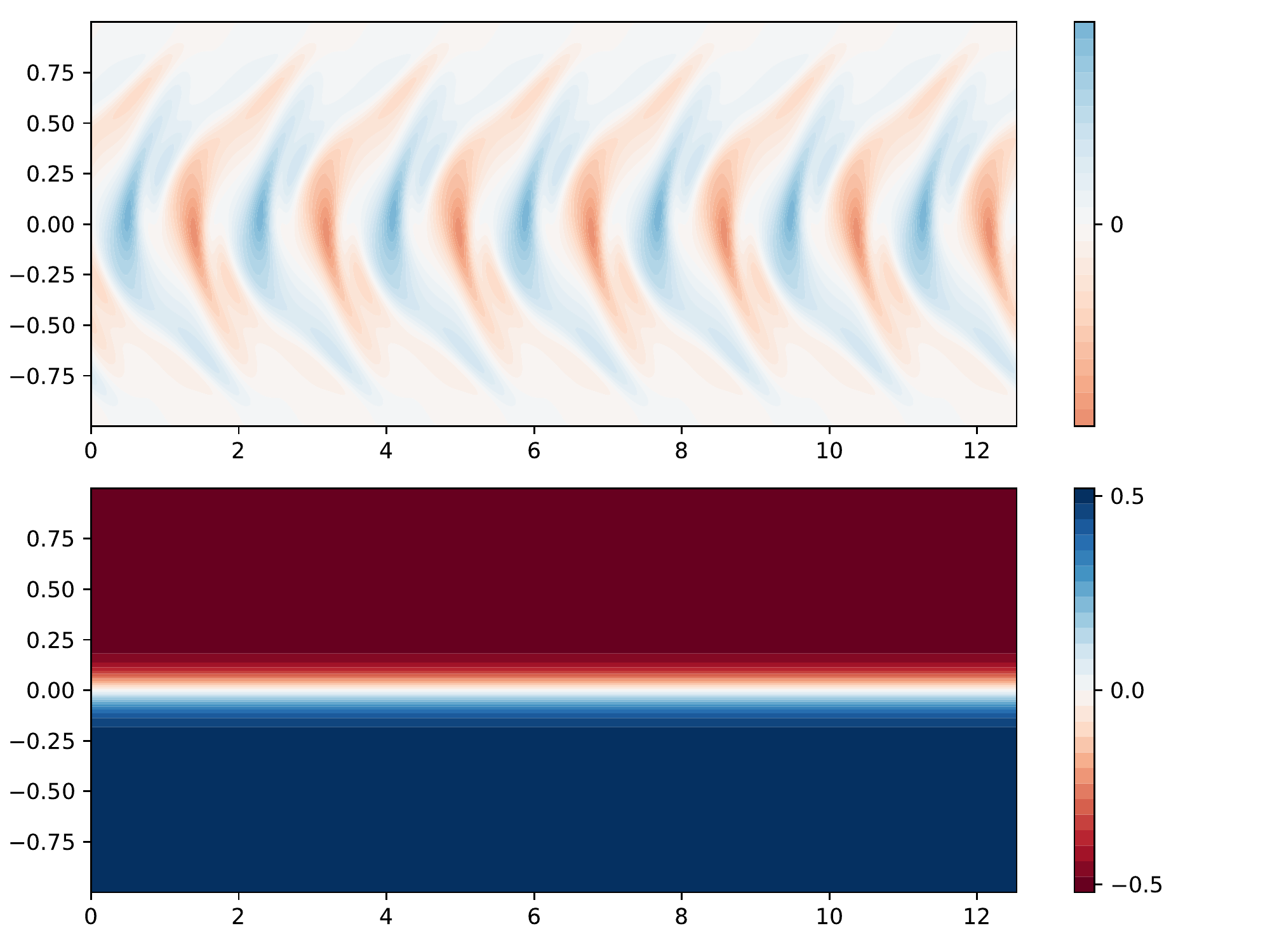}
    \caption{$b_0$ (bottom) and $\nabla \times \boldsymbol{u}_0$ (top).}
    \end{subfigure}
    \hfill
    \begin{subfigure}[b]{0.495\textwidth}
     \centering
     \includegraphics[width=0.9\textwidth]{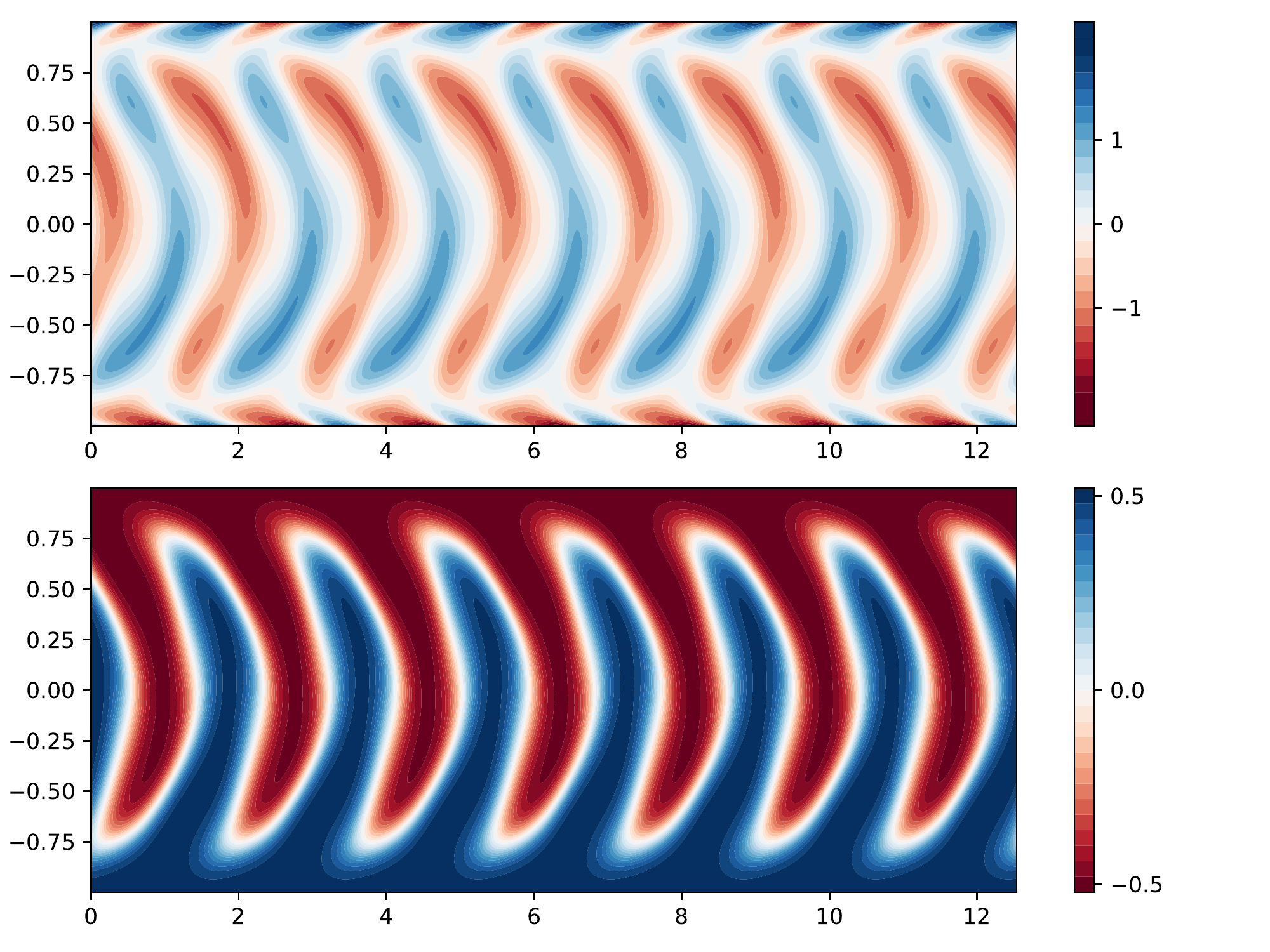}
     \caption{$b_T$ (bottom) and $\nabla \times \boldsymbol{u}_T$ (top).}
    \end{subfigure}
     \caption{The optimal (a) initial velocity perturbation (given in terms of the vorticity $\nabla \times \boldsymbol{u}_0$) and (b) the resulting deviation density and perturbation velocity fields at $T=5$ for $Ri=0.05$. To be compared with figure 2 of \cite{marcotte2018optimal}. This optimum was calculated using a Wolfe based line-search and conjugate gradient method corresponding to \autoref{table:Optimisation_Test_Mixing}(f).}
     \label{fig:DAL_CONV_Mixing_Result}
\end{figure}

\autoref{fig:DAL_CONV_Mixing_Result}(a) shows the sharp initial density deviation $b_0$ and optimal velocity perturbation (here plotted as the vorticity $\nabla \times \boldsymbol{u}_0$), which as shown \autoref{fig:DAL_CONV_Mixing_Result}(b) allows the mixing process to unfold via Taylor dispersion \cite{marcotte2018optimal}. These figures, computed by solving \eqref{eq:Mixing_Objective} using the discrete gradient and conjugate gradient method accompanied by a Wolfe line-search (\autoref{table:Optimisation_Test_Mixing}(f)), also demonstrate that our implementation faithfully reproduces the results of \cite{marcotte2018optimal}. 

\subsubsection{Discrete vs. continuous adjoint}

With our code's validity established we now compare the performance of the discrete and continuous gradient when optimising for the perturbation $\boldsymbol{u}_0$. The convergence of these runs reported in \autoref{table:Optimisation_Test_Mixing}, establish that only the discrete gradient can satisfy the imposed convergence criteria. Although the conjugate gradient method using the continuous gradient works well initially, it subsequently stalls due to a poor gradient estimate. While both gradient estimates choose a dominant wavelength $k_x=7$ the discrete gradient consistently achieves a slightly better optimum than the continuous gradient.

\begin{table}[h!]
\caption{\label{table:Optimisation_Test_Mixing} Test summary and numerical cost of reaching a stationary point using different descent routines as shown in figures \ref{fig:DAL_CONV_Mixing_Result} and \ref{fig:DAL_CONV_MIX}. We set $\alpha_{max}=100$ and terminate optimisation when $r^i_k \leq$1e-06 or iterations $>200$. Star $*$ denotes a stalled run. Rows (a),(b) report the \texttt{continuous residual} and (c)-(f) the discrete residual.}
\centering
\begin{tabular}{ c | c|  c | c| c | c | c}
 \hline
Routine & Gradient & Line-search & evals $\hat{J}(\boldsymbol{u},b)$ & evals $\frac{\delta \mathcal{L}}{\delta \boldsymbol{u}_0}$ & $r_k$ & $\hat{J}(\boldsymbol{u},b)$  \\
\hline
(a) SD & Continuous & Armijo        & 209  & 200 & \texttt{4.9943e-04} & 1.2033e-02\\
\hline
(b) CG & Continuous & Wolfe I \& II & 65   & 61  & *\texttt{5.5593e-04} & *1.2032e-02 \\
\hline
(c) SD & Discrete   & Armijo        & 209  & 200 & 1.0852e-04 & 1.2030e-02\\
\hline
(d) SD & Discrete   & Wolfe I \& II & 258  & 248 & 7.5664e-07 & 1.2029e-02\\
\hline
(e) CG & Discrete   & Armijo        & 209  & 200 & 9.5897e-07 & 1.2030e-02\\
\hline
(f) CG & Discrete   & Wolfe I \& II & 123  & 116 & 7.6475e-07 & 1.2029e-02\\
\hline
\end{tabular}
\end{table}

\subsubsection{Line-search algorithms}

\begin{figure}[ht]
    \centering
    \begin{subfigure}{0.325\textwidth}
     \centering
     \includegraphics[scale=0.28]{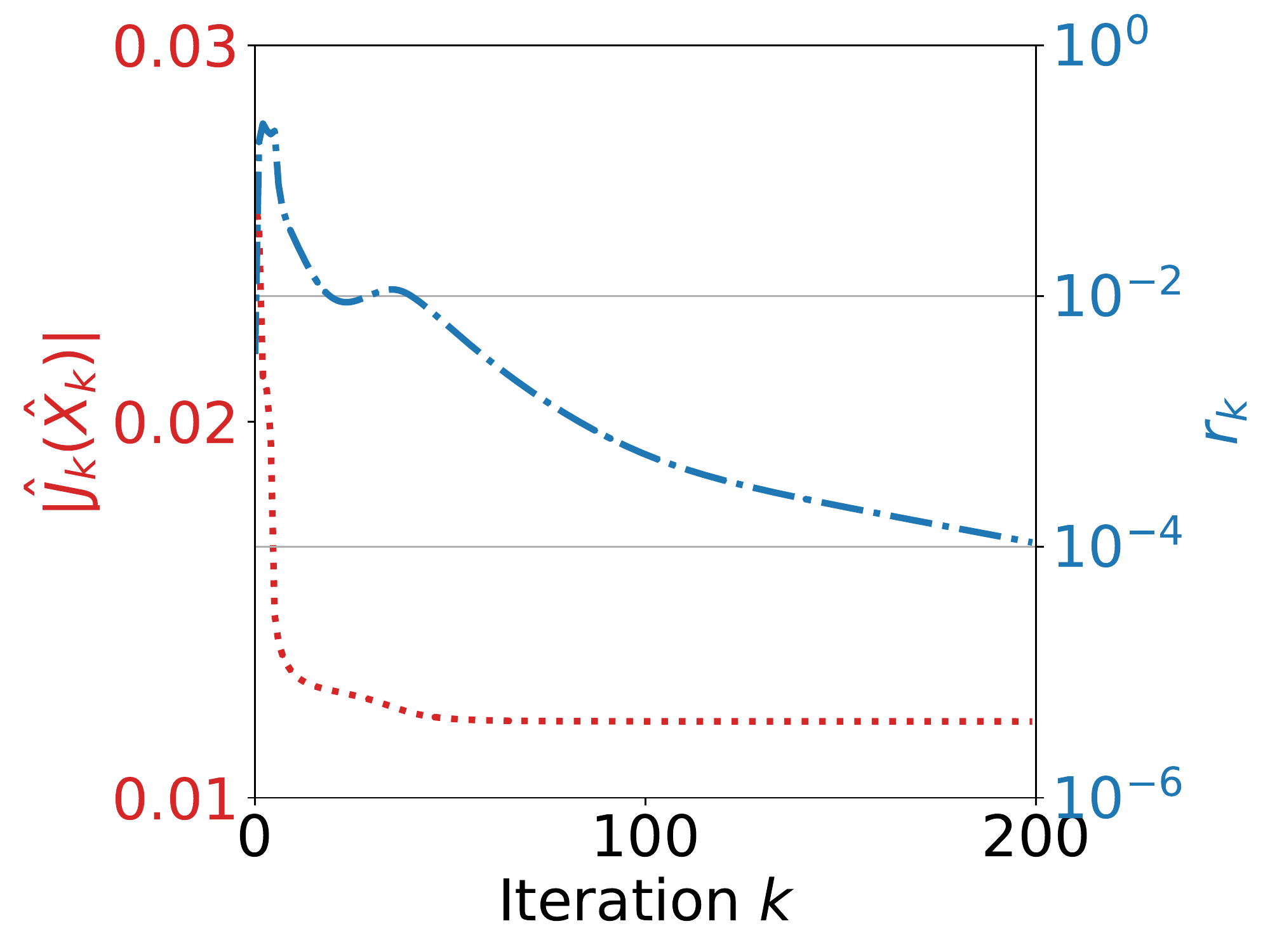}
     \caption*{(c) SD (Armijo)}
    \end{subfigure}
    \begin{subfigure}{0.325\textwidth}
     \centering
     \includegraphics[scale=0.28]{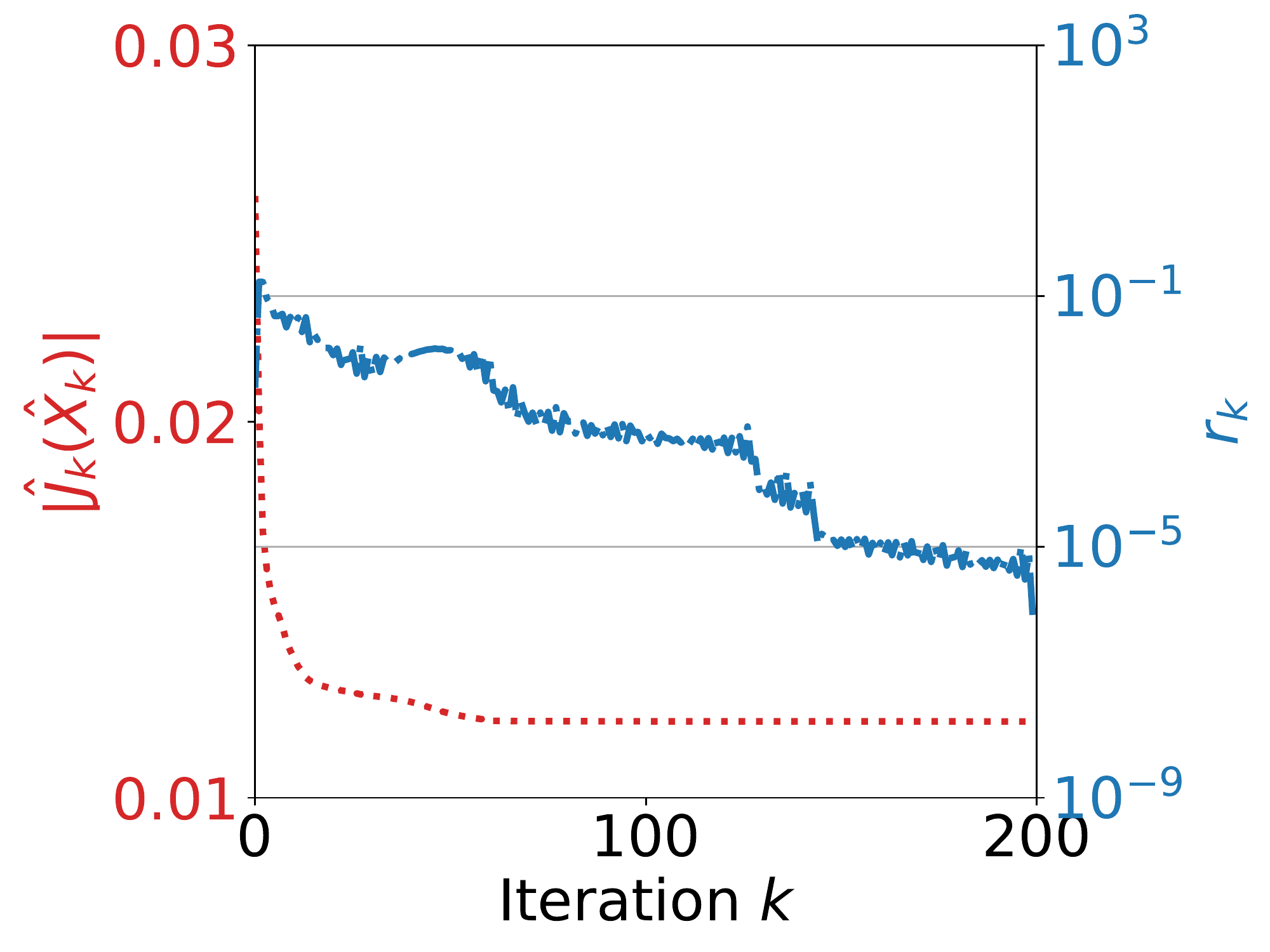}
     \caption*{(d) SD (Wolfe I \& II)}
    \end{subfigure}
    \begin{subfigure}{0.325\textwidth}
     \centering
     \includegraphics[scale=0.28]{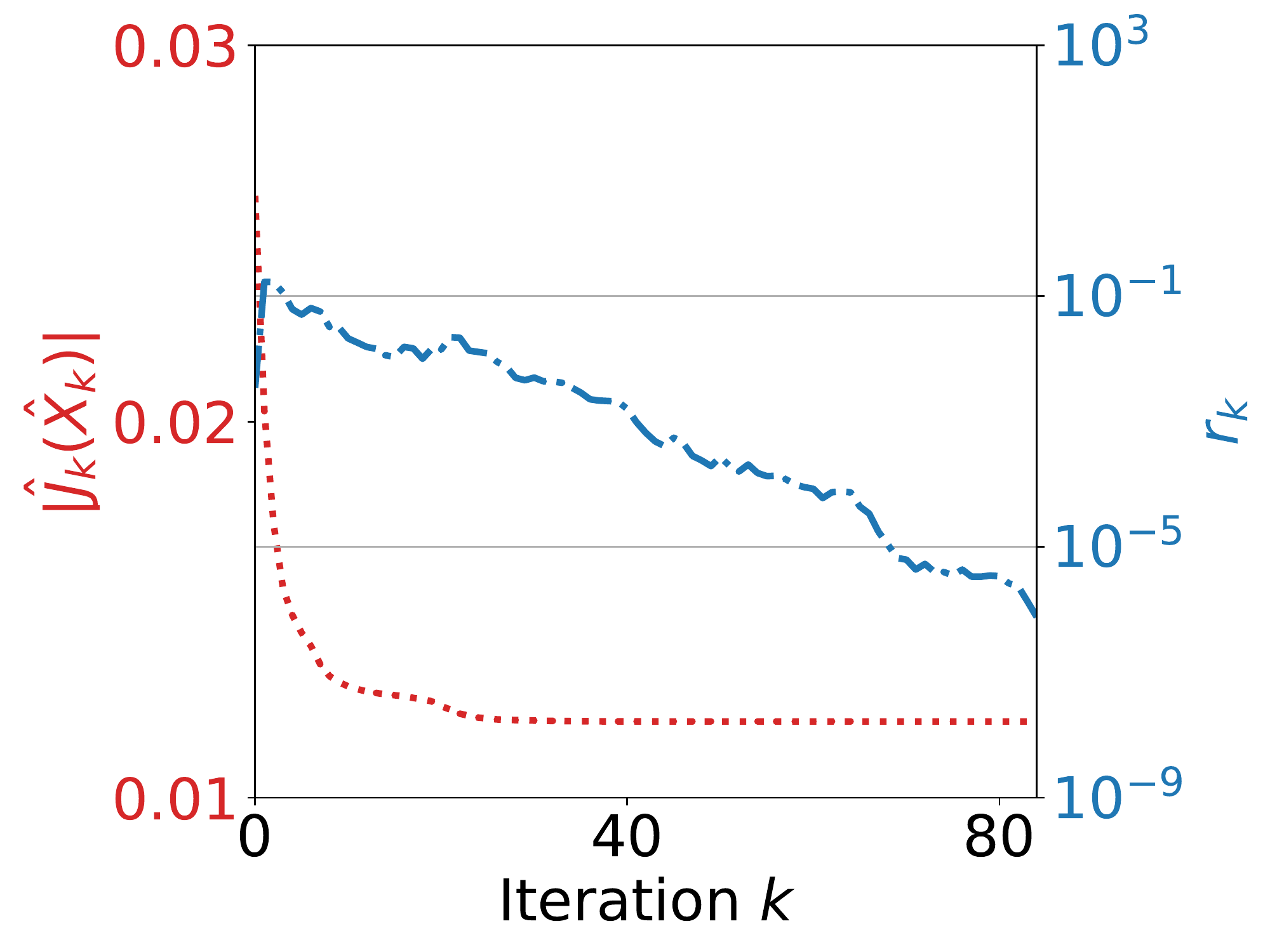}
     \caption*{(f) CG (Wolfe I \& II)}
    \end{subfigure}
     \caption{Convergence of the optimisation runs in rows (c)-(f) of \autoref{table:Optimisation_Test_Mixing}, in terms of the cost function $J$ (dotted, red) and gradient residual $r$ (chained, blue), to the perturbation $\boldsymbol{u}_0$ when initiated with random noise. The steepest descent (SD) methods (c),(d) converge sub-linearly while the conjugate-gradient (CG) method (f) at least linearly, thus demanding fewer iterations.}
     \label{fig:DAL_CONV_MIX}
\end{figure}

As shown in \autoref{fig:DAL_CONV_MIX} and \autoref{table:Optimisation_Test_Mixing} the discrete gradient allows us to use the conjugate gradient algorithm and obtain fast convergence. Combining the discrete gradient with either a conjugate gradient method (e) or a strong Wolfe line-search (f) the imposed convergence criteria can also be satisfied, albeit less efficiently. Comparing row (e) with (f) and row (c) with (d), it is also apparent that the strong Wolfe line-search greatly influences convergence despite its greater cost.

\subsection{Maximum step-size}

For each problem considered we have prescribed a hyper-parameter $\alpha_{max}$. For an Armijo line-search it is the initial step-size $\alpha_0 = \alpha_{max}$ from which back-tracking begins, while in a strong Wolfe line-search this constitutes the maximum admissible step-size. By reducing or increasing the number of iterations required to achieve convergence, this parameter was found to have a substantial impact on performance, in all problems considered. Typically as was the case in the PCA and Swift-Hohenberg problems setting $\alpha_{max} \approx 1$ works well. In the kinematic dynamo and optimal mixing problems however we found that the range of step sizes considered by the line-search must be larger, particularly when using the strong Wolfe-conditions $(\alpha_{min},\alpha_{max}) = (10^{-6},10^2)$; although steepest descent was also found to converge in 20-40 fewer iterations when a large maximum/initial step-size was chosen for these problems. We understand this behaviour to be a consequence of $\hat{J}(\hat{\boldsymbol{X}}_{k+1})$ being (informally speaking) \emph{poorly bounded} during initial iterations due to the unfavourable initial dynamo velocity field $\boldsymbol{U}$ or mixing perturbation velocity field $\boldsymbol{u}$. Consequently this requires selecting a larger than usual $\alpha_k$ for the Armijo condition \eqref{eq:Armijo} to become active and for \eqref{eq:Wolfe_Conditions_I_II} be easily satisfied. A further discussion of this issue is given in the introduction of \cite{more1994line}.


\section{Conclusion}\label{sec:Conclusion}

In this paper we have focused on inconsistencies arising from the gradient estimation and application of equality constraints in the numerical direct adjoint-based control procedure \cite{Kerswell2018}; to show why these issues occur and how they can be resolved on practical examples. Deriving the discrete adjoint equations of time-dependent PDE constraints for (preconditioned) multistep schemes \cite{Ascher1995}, and considering three example problems we demonstrated in \autoref{sec:Discrete_Adjoint} a consistent gradient approximation, readily obtainable with minor code modifications. In \autoref{sec:Linesearch_Algorithim} we recalled following \cite{boumal2020introduction,sato2021riemannian_book} how an equality constrained problem can be solved as an unconstrained problem on a spherical manifold. Considering three examples, we numerically demonstrate that this procedure, available as \texttt{SphereManOpt} \cite{SphereManOpt2022}, is capable of giving robust linear to super-linear convergence to stationary points with small residual error when both nonlinear PDE and multiple norm constraints are present. Useful extensions of this work would consider the discrete adjoint of Runge-Kutta schemes, preconditioning schemes for problems with multiple norm constraints and exploiting additional geometric structures of the optimisation problems considered. Further developments of \texttt{SphereManOpt} would include $L_p$ norms and the integration of an optimal checkpointing strategy \cite{griewank2000algorithm}.

\appendix

\section{Preconditioned first order scheme}\label{App:Preconditioned_First_Order_Scheme}

Departing from \eqref{eq:TimeStep_Notation_multistep} we write a general \emph{s-step} multistep IMEX method as
\begin{equation}
    \big(a_0 \mathcal{M}_p + b_0 \mathcal{L}_p \big) \hat{\boldsymbol{X}}^n_p = \sum_{j=1}^2 c_j \hat{\mathcal{F}}^{n-j}_p - a_j \mathcal{M}_p \hat{\boldsymbol{X}}^{n-j}_p - b_j \mathcal{L}_p \hat{\boldsymbol{X}}^{n-j}_p,
    \label{eq:Multistep_Cheby_Precond}
\end{equation}
where $\hat{\boldsymbol{X}} = T \boldsymbol{X}$ denotes the Chebyshev transform of the first kind and the matrix operators $\mathcal{M},\mathcal{L}$ are assumed to be in spectral space such that $\hat{\boldsymbol{X}}$ corresponds to a vector of Fourier modes and Chebyshev polynomial coefficients. For the purposes of parallelisation, the previous equation is separated into pencils i.e. equations separable in Fourier modes such that only different Chebyshev modes couple. In 2D for example the number of pencils corresponds exactly to the number of Fourier modes divided by two, owning to a complex to real conversion. We thus consider the solution procedure for individual pencils as indexed by subscript $p$ in \eqref{eq:Multistep_Cheby_Precond}, where only the right hand side couples different Fourier/transverse modes. 

To efficiently solve \eqref{eq:Multistep_Cheby_Precond} {\sc Dedalus} employs preconditioners $P_p^L, P^R$ with the former depending on the pencil considered. By using: a first order formulation, converting from Chebyshev polynomials of the first kind $T_n(z)$ to Chebyshev polynomials of the second kind $U_n(z)$, performing Dirichlet recombination and grouping by Chebyshev modes of the second kind rather than variables, all left hand side (LHS) matrices are rendered sparse \cite{Burns2020}. In practice these preconditioners are employed as
\begin{equation}
     \underbrace{ \big(a_0 \bar{\mathcal{M}_p} + b_0 \bar{\mathcal{L}_p} \big) }_{A_p} \underbrace{ (P^R)^{-1} \hat{\boldsymbol{X}}^n_p }_{\hat{\boldsymbol{Y}}^n_p} = \underbrace{ P_p^L \sum_{j=1}^s c_j \hat{\mathcal{F}}^{n-j}_p - a_j \mathcal{M}_p \hat{\boldsymbol{X}}^{n-j}_p - b_j \mathcal{L}_p \hat{\boldsymbol{X}}^{n-j}_p }_{B_p},
\end{equation}
with $\bar{\mathcal{M}_p} = P_p^L\mathcal{M}_p P^R$ to allow for a sparse inversion of the banded system
\begin{equation}
\hat{\boldsymbol{Y}}^n_p = A^{-1}_p B_p, \quad \hat{\boldsymbol{X}}^n_p = (P^R) \hat{\boldsymbol{Y}}^n_p.
\end{equation}
Similarly $\hat{\boldsymbol{X}}^n_p$ is easily recovered by left multiplication using $P^R$. Using the recurrence relation $2 T_n(z) = U_n(z) - U_{n-2}(z)$ we have a relation between the two sets of polynomials. To convert between them we use the left preconditioner $P^L_p$. The above constitutes the solution method followed for the optimal mixing problem \eqref{eq:Mixing_Lagrangian} and what follows the method for computing its adjoint equations.

To solve \eqref{eq:Disc_Euler_Lagrange_CNAB1_0} to \eqref{eq:Disc_Euler_Lagrange_CNAB1} using this formulation, we retain only the first order coefficients denoted by a tilde, multiply by the corresponding left preconditioner $P_p^L$, insert the right preconditioner and its inverse so as to form $A_p$ and finally form their corresponding adjoint equations at indices $n=N$, $1 \leq n \leq N-1$ and $n=0$ respectively
\begin{eqnarray}
 \big(a_0 \bar{\mathcal{M}^{\dagger}_p} + b_0 \bar{\mathcal{L}^{\dagger}_p} \big) \hat{\boldsymbol{q}}^N_p = (P^R)^{\dagger} \frac{\partial J }{\partial \hat{\boldsymbol{X}}_N}, \label{eq:Pre_Disc_Euler_Lagrange_CNAB1_0} \\
 \big(a_0 \bar{\mathcal{M}^{\dagger}_p} + b_0 \bar{\mathcal{L}^{\dagger}_p} \big) \hat{\boldsymbol{q}}^n_p = -\bigg[ \big( \tilde{a}_1 \bar{\mathcal{M}^{\dagger}_p} + \tilde{b}_1 \bar{\mathcal{L}^{\dagger}_p} \big)  - (P^R)^{\dagger} \tilde{c}_1 \frac{ \partial F }{\partial \boldsymbol{X}_n}^{\dagger} (P^L_p)^{\dagger} \bigg] \hat{\boldsymbol{q}}^{n+1}_p +
(P^R)^{\dagger} \frac{\partial J}{\partial \hat{\boldsymbol{X}}_n} \\
  \frac{\partial \mathcal{L}}{\partial \boldsymbol{X}(t=0)} \equiv \hat{\boldsymbol{q}}^0_p = -\bigg[ (\tilde{a}_1 \mathcal{M}^{\dagger} + \tilde{b}_1 \mathcal{L}^{\dagger} ) - \tilde{c}_1 \frac{ \partial F }{\partial \boldsymbol{X}_0}^{\dagger} \bigg] (P^L_p)^{\dagger} \boldsymbol{q}_1 + \frac{\partial J}{\partial \hat{\boldsymbol{X}}_0}. \label{eq:Pre_Disc_Euler_Lagrange_CNAB1}
\end{eqnarray} 
The derivation shows that when implemented in Dedalus or an alternative code that avails of preconditioning matrices an additional source of error arises from the non-commutative nature of the Hermitian adjoint operation. To compute the variation of the objective function requires computing the adjoint of the Chebyshev transforms to see why this arises we consider the $\mathcal{O}(\Delta t)$ accurate time integration of the objective function
\begin{equation}
    J(\hat{\boldsymbol{X}}) = \sum_{n=0}^N \frac{1}{2V} \langle T^{-1} \hat{\boldsymbol{X}}_n, W T^{-1} \hat{\boldsymbol{X}}_n \rangle \Delta t,
\end{equation}
where $W=W^{\dagger}$. Taking variations of the above we obtain 
\begin{equation}
    \langle \delta \hat{\boldsymbol{X}_n}, \frac{\partial J}{\partial \hat{\boldsymbol{X}}_n} \rangle = \langle  \delta \hat{\boldsymbol{X}}_n, T^{-\dagger} W T^{-1} \hat{\boldsymbol{X}}_n  \frac{\Delta t}{V} \rangle.
\end{equation}

\section{Discrete adjoint of the Swift-Hohenberg problem}

Writing the optimisation problem \eqref{eq:SH23_Objective} as a minimisation problem in terms of the Lagrangian
\begin{equation}
    \mathcal{L}(u,u^{\dag}) = - \int_{t=0}^{T} \langle u(x,t), u(x,t) \rangle dt - \int_{t=0}^{T} \langle u^{\dag}, \mathcal{M} \partial_t u + \mathcal{L} u - \mathcal{F}(u,t) \rangle dt,
    \label{eq:SH23_Lagrangian}
\end{equation}
where $\langle f, g \rangle = \int_x fg \; dx, \; \mathcal{M} = \mathbb{I}, \mathcal{L} = (1+\partial^2_x)^2 -a$ and $\mathcal{F}(u,t) = 1.8u^2 - u^3$. The adjoint equations and compatibility conditions corresponding to \eqref{eq:SH23_Lagrangian} give
\begin{equation}
    -\partial_t u^{\dag} + (1+\partial^2_x)^2u^{\dag} -au^{\dag} = (3.6u - 3 u^2) u^{\dag} - 2u, \quad\quad u^{\dag}(x,T) = 0.
    \label{eq:SH23_Lagrangian_Adjoint}
\end{equation}
Letting $\boldsymbol{u}(t)$ denote the spatial discretisation of $u(x,t)$ and $\tilde{\boldsymbol{u}} = \left( \boldsymbol{u}_0, \cdots, \boldsymbol{u}_N \right)$ its temporal discretisation such that $t_n = n \Delta t, 0 \leq n \leq N$, its $\mathcal{O}(\Delta t)$ accurate discrete counterpart is given by
\begin{eqnarray}
( \tilde{a}_0 \mathcal{M} + \tilde{b}_0 \mathcal{L} ) \boldsymbol{u}^{\dag}_N = -2 \boldsymbol{u}_N,
\\
( \tilde{a}_0 \mathcal{M}   + \tilde{b}_0 \mathcal{L} )\boldsymbol{u}^{\dag}_n + ( \tilde{a}_1  \mathcal{M} + \tilde{b}_1 \mathcal{L} )\boldsymbol{u}^{\dag}_{n+1} - \tilde{c}_1 \frac{ \partial \mathcal{F}}{\partial \boldsymbol{u}}^{\dagger} |_n \boldsymbol{u}^{\dag}_{n+1} = -2 \boldsymbol{u}_n,
\\
\frac{\boldsymbol{u}^{\dag}_0}{\Delta t} - (\tilde{a}_1 \mathcal{M} + \tilde{b}_1 \mathcal{L} ) \boldsymbol{u}^{\dag}_1 - \tilde{c}_1 \frac{ \partial \mathcal{F}}{\partial \boldsymbol{u}}^{\dagger} |_0 \boldsymbol{u}^{\dag}_1 = -2 \boldsymbol{u}_0,
\label{eq:SH23_Disc_Euler_Lagrange_CNAB1}
\end{eqnarray}
for $n=N,1 \leq n \leq N-1,n=0$ respectively, where $\frac{ \partial \mathcal{F}}{\partial \boldsymbol{u}}^{\dagger} |_n \boldsymbol{u}^{\dag}_{n+1} = (3.6\boldsymbol{u}_n - 3 \boldsymbol{u}^2_n ) \boldsymbol{u}^{\dag}_{n+1}$.

\section{Discrete adjoint of the kinematic dynamo problem}\label{Dynamo_adjoint}

The continuous Lagrangian corresponding to \eqref{eq:Max_kinematic} is given by
\begin{equation}
\begin{split}
    \mathcal{L} &= \langle \boldsymbol{B}^2_T \rangle - \int_{0}^{T} \langle \boldsymbol{B}^{\dagger}, \frac{\partial \boldsymbol{B} }{\partial t} - \nabla \times \big( \boldsymbol{U} \times \boldsymbol{B} \big) - \nabla \Pi -  Rm^{-1} \nabla^2 \boldsymbol{B} \rangle dt \\
& - \int_{0}^{T} \langle \Pi^{\dagger}, ( \nabla \cdot \boldsymbol{B}) \rangle dt - \int_{0}^{T} \langle P^{\dagger}, (\nabla \cdot \boldsymbol{U} ) \rangle dt,
\end{split}
\label{eq:Max_Kinematic_Lagrangian}
\end{equation}
where $\boldsymbol{B}^{\dagger},\Pi^{\dagger},P^{\dagger}$ are the adjoint variables enforcing the constraint equations for the adjoint variables. Taking variations with respect to the primal variables we obtain
\begin{equation}
\begin{split}
    \frac{ \delta \mathcal{L} }{\delta \boldsymbol{B} }   & = \frac{\partial \boldsymbol{B}^{\dagger}}{\partial t} + Rm^{-1} \nabla^2 \boldsymbol{B}^{\dagger} + \nabla \Pi^{\dagger} + \big( \nabla \times \boldsymbol{B}^{\dagger} \big) \times \boldsymbol{U} = 0 \\
    \frac{ \delta \mathcal{L} }{\delta \boldsymbol{\Pi} }  & = \nabla \cdot \boldsymbol{B}^{\dagger} = 0, \\
    \frac{ \delta \mathcal{L} }{\delta \boldsymbol{B}_0 } &= \boldsymbol{B}^{\dagger}_0, \\
    \frac{ \delta \mathcal{L} }{\delta \boldsymbol{B}_T } & = 2 \boldsymbol{B}_T - \boldsymbol{B}^{\dagger}_T = 0, \\
    \frac{ \delta \mathcal{L} }{\delta \boldsymbol{U} }   &= \int^T_0 \big[ \boldsymbol{B} \times (\nabla \times \boldsymbol{B}^{\dagger}) + \nabla P^{\dagger} \big] dt,
\end{split}
\label{eq:Euler_Lagrange_Kinematic_Adjoint}
\end{equation}
for which the last line can be replaced by
\begin{equation}
\frac{\partial }{ \partial t} \bigg( \frac{ \delta \mathcal{L} }{\delta \boldsymbol{U} } \bigg)  = \boldsymbol{B} \times (\nabla \times \boldsymbol{B}^{\dagger}) + \nabla P^{\dagger}, \quad \nabla \cdot \frac{ \delta \mathcal{L} }{\delta \boldsymbol{U} } =0,    
\end{equation}
to ensure a divergence free gradient. The discrete counterpart of \eqref{eq:Max_Kinematic_Lagrangian} is given by
\begin{equation} 
\begin{split}
    \mathcal{L} &= \langle \boldsymbol{X}_N, \mathcal{M}\boldsymbol{X}_N \rangle  - \langle \boldsymbol{q}_0, \boldsymbol{X}_0 - \boldsymbol{X}(\boldsymbol{x},t=0) \rangle \\
& -\sum_{n=0}^{N-1} \langle \boldsymbol{q}_{n+1}, \mathcal{M}(\tilde{a}_0\boldsymbol{X}_{n+1} + \tilde{a}_1 \boldsymbol{X}_n) + \mathcal{L}(\tilde{b}_0\boldsymbol{X}_{n+1} + \tilde{b}_1\boldsymbol{X}_n) - \tilde{c}_1 \underbrace{\left( \nabla \times (\boldsymbol{U} \times \boldsymbol{B}_n) , \; 0 \right)}_{\mathcal{F}(\boldsymbol{X}_n)} \rangle \Delta t,
\end{split}
\label{eq:Max_Kinematic_Lagrangian_Discrete}
\end{equation}
where the vectors and operators are given by
\begin{equation}
 \boldsymbol{X}_n = (\boldsymbol{B}, \Pi)^T_n, \quad \boldsymbol{q}_n = (\boldsymbol{B}^{\dagger}, \Pi^{\dagger})_n, \quad \mathcal{M} = diag(\mathbb{\boldsymbol{I}},0), \quad
\mathcal{L} = \left(
\begin{array}{*{2}c}
-Rm^{-1} \boldsymbol{\Delta} & \nabla \\
{\nabla}^T & 0
\end{array} \right).
\end{equation}
Taking variations with respect to $\boldsymbol{X}_n$ we recover the discrete adjoint equations
\begin{eqnarray}
( \tilde{a}_0 \mathcal{M} + \tilde{b}_0 \mathcal{L} ) \boldsymbol{q}_N = \Delta t^{-1} 2 \mathcal{M}  \boldsymbol{X}_N, \label{eq:Disc_Kinematic_Euler_Lagrange_CNAB1}
\\
( \tilde{a}_0 \mathcal{M} + \tilde{b}_0 \mathcal{L} ) \boldsymbol{q}_n + ( \tilde{a}_1 \mathcal{M} + \tilde{b}_1 \mathcal{L} ) \boldsymbol{q}_{n+1} - \tilde{c}_1 \frac{ \partial \mathcal{F} }{\partial \boldsymbol{X}_n}^{\dagger}\boldsymbol{q}_{n+1} = 0,
\\
\boldsymbol{q}_0 \Delta t^{-1} - (\tilde{a}_1 \mathcal{M} + \tilde{b}_1 \mathcal{L} ) \boldsymbol{q}_1 - \tilde{c}_1 \frac{ \partial \mathcal{F} }{\partial \boldsymbol{X}_n}^{\dagger} \boldsymbol{q}_1= 0,
\end{eqnarray}
\noindent for $n=N,N-1 \geq n \geq 1,n=0$ respectively, where
\begin{equation}
    \frac{ \partial \mathcal{F} }{\partial \boldsymbol{X}_n}^{\dagger} \boldsymbol{q}_{n+1} = \left( ( \nabla \times \boldsymbol{B}^{\dag}_{n+1} ) \times \boldsymbol{U} , \; 0 \right).
\end{equation}
In the case where $\boldsymbol{U}$ is allowed to be updated, we take variations with respect to $\boldsymbol{U}$ directly in \eqref{eq:Max_Kinematic_Lagrangian_Discrete}  in order to obtain the additional equation
\begin{equation}
    \frac{ \delta \mathcal{L}(\boldsymbol{X}_N,\boldsymbol{q}_N) }{\delta \boldsymbol{U} } = \sum_{n=0}^{N-1} \big[ -\tilde{c}_1 \boldsymbol{B}_n \times (\nabla \times \boldsymbol{B}^{\dag}_{n+1}) + \nabla P^{\dagger}_{n+1} \big]\Delta t,
\end{equation}
which is time-stepped by re-writing the above as
\begin{equation}
    ( \tilde{a}_0 \mathcal{M} + \tilde{b}_0 \mathcal{L} ) \boldsymbol{w}_n + ( \tilde{a}_1 \mathcal{M} + \tilde{b}_1 \mathcal{L} ) \boldsymbol{w}_{n+1} = 
    -\tilde{c}_1 \left( \boldsymbol{B}_n \times (\nabla \times \boldsymbol{B}^{\dag}_{n+1}),  \;  0 \right),
    \label{eq:Disc_Kinematic_Euler_Velocity}
\end{equation}
where $\boldsymbol{w}_n = (\frac{\delta \mathcal{L}}{\delta \boldsymbol{U}}, \
P^{\dagger})_n$ and $\mathcal{L}$ is as above but with $Rm^{-1} =0$ such that $\nabla P^{\dagger}$ is used to enforce $\nabla \cdot \boldsymbol{\frac{\delta \mathcal{L}}{\delta \boldsymbol{U}}}=0$. By enforcing both divergence free constraints at all times using the Lagrange multipliers $\Pi, \Pi^{\dag}, P^{\dag}$ we eliminates the need for projection steps including Chorin's method which can complicate gradient computations \cite{de2012efficient}. In all cases we apply the gauge condition that the zero'th Fourier mode of each field $f_{k_x=0,k_y=0,k_z=0} = 0$.

\section{Discrete adjoint of the optimal mixing problem}\label{sec:Mixing_Adjoint}

The continuous Lagrangian corresponding to \eqref{eq:Mixing_Objective} is given by
\begin{equation}
\begin{split}
     \mathcal{L} &= \mathcal{J}(\boldsymbol{u},b) - \int^T_0 \langle \boldsymbol{u}^{\dag}, \frac{\partial \boldsymbol{u}}{\partial t} + (\boldsymbol{U}_0 \cdot \nabla) \boldsymbol{u} + (\boldsymbol{u} \cdot \nabla)[\boldsymbol{U}_0 + \boldsymbol{u}] - \nabla p + Ri b \hat{\boldsymbol{z}} - \frac{1}{Re} \nabla^2 \boldsymbol{u} \rangle \; dt \\
                 & - \int^T_0 \langle b^{\dag}, \frac{\partial b}{\partial t} + ([\boldsymbol{U}_0 + \boldsymbol{u}] \cdot \nabla) b - \frac{1}{Pe} \nabla^2 b \rangle \; dt - \int^T_0 \langle p^{\dag}, \nabla \cdot \boldsymbol{u} \rangle \; dt,
\end{split}
\label{eq:Mixing_Lagrangian}    
\end{equation}
where $\boldsymbol{u}^{\dag},p^{\dag},b^{\dag}$ are the adjoint variables enforcing the constraint equations for the dual variables. Taking variations with respect to the primal variables we obtain the adjoint equations 
\begin{equation}
\begin{split}
&\frac{\partial \boldsymbol{u}^{\dag}}{\partial t} - \frac{1}{Re} \nabla^2 \boldsymbol{u}^{\dag} - \nabla p^{\dag} -(\boldsymbol{U}_0 \cdot \nabla) \boldsymbol{u}^{\dag} + \boldsymbol{u}^{\dag} \cdot (\nabla \boldsymbol{U}_0 )^T = (\boldsymbol{u} \cdot \nabla) \boldsymbol{u}^{\dag} - \boldsymbol{u}^{\dag} \cdot (\nabla \boldsymbol{u})^T - b^{\dag} \nabla b, \\ 
& \frac{\partial b^{\dag} }{\partial t} - \frac{1}{Pe} \nabla^2 b^{\dag} + Ri \boldsymbol{u}^{\dag} \cdot \hat{\boldsymbol{z}} -(\boldsymbol{U}_0 \cdot \nabla) b^{\dag} = (\boldsymbol{u} \cdot \nabla) b^{\dag}, \quad \nabla \cdot \boldsymbol{u}^{\dag} = 0, 
\end{split}
\label{eq:Euler_Lagrange_Mixing_Adjoint}   
\end{equation}
and compatibility conditions
\begin{equation}
    \boldsymbol{u}^{\dag}(\boldsymbol{x},t=T) = 0, \quad\quad b^{\dag}(\boldsymbol{x},t=T) = - \nabla^{-2} b(\boldsymbol{x},T), 
\end{equation}
where $b^{\dag}$ follows from solving the linear boundary value problem 
\begin{equation}
\nabla^2 \psi= b, \quad \hbox{s.t.} \quad \partial_z \psi(x,z=\pm1) =0, \; \int_{V} \psi \; dV =0.
\label{eq:lbvp}
\end{equation}
Whilst the discrete adjoint for these equations can be computed in a similar manner following \ref{App:Preconditioned_First_Order_Scheme}, the discrete implementation of this mix-norm cost functional, and its adjoint, merits special attention. The mix norm cost functional takes the form
\begin{equation}
    \mathcal{J}= \langle \nabla\psi,W \nabla\psi\rangle,
\end{equation}
where $W$ is an appropriate weight matrix and the symbol $\nabla$ is now taken to mean the discretised gradient operator. With this in consideration, the cost functional can be rewritten in terms of $b$ as 
\begin{equation}
    \mathcal{J}=\langle \nabla\nabla^{-2} b,W \nabla\nabla^{-2}b\rangle,
\end{equation}
giving the final-time condition
\begin{equation}
    b^\dagger(\boldsymbol{x},T) =(\nabla^{-2})^{\dagger} \nabla^{\dagger} W \nabla \psi. 
    \label{equ:rho_final_time_disc}
\end{equation} 
Note that in deriving (\ref{equ:rho_final_time_disc}) we have assumed that $W^{\dagger}=W$. Numerically we decompose $\nabla$ as $(\partial_x,\partial_z)^T$ where $\partial_x$ and $\partial_z$ are the discrete $x$ and $z$ derivatives and denote by $(\partial_x^\dagger,\partial_z^\dagger)$ their discrete-adjoint counterpart. The discrete adjoint for $\nabla^{-2}$ is found directly from the {\sc Dedalus} implementation of \eqref{eq:lbvp}. In this manner, we find $b^\dagger(\boldsymbol{x},T)$ by solving $(\nabla^2)^{\dagger} b^\dagger(\boldsymbol{x},T)=\nabla^{\dagger}\mathbf{W}^{\dagger}\nabla \psi$.

\section*{Acknowledgements}

The authors wish to thank Ashley P. Willis for his careful reading and recommendations for improving this manuscript. P. M. M. and F. M. acknowledge support from the French program “T-ERC” from Agence Nationale de la Recherche (ANR) under grant agreement ANR-19-ERC7-0008). C. S. S. acknowledges partial support from a grant from the Simons Foundation (Grant No. 662962, GF). He would also like to acknowledge support of funding from the European Union Horizon 2020 research and innovation programme (grant agreement no. D5S-DLV-786780). This work was also supported by the French government, through the UCAJEDI Investments in the Future project managed by the ANR under grant agreement ANR-15-IDEX-0001. The authors are grateful to the OPAL infrastructure from Universit\'e C\^ote d’Azur, Universit\'e C\^ote d’Azur’s Center for High-Performance Computing computer facilities for providing resources and support.

\bibliographystyle{plain}

\bibliography{IOPLaTeXGuidelines}

\end{document}